\documentclass[final]{article}
\usepackage{graphicx,multirow}
\usepackage{newproof}

\begin{document}

\title{\large \bf Reversible circuit synthesis using a cycle-based approach}
\author{\normalsize Mehdi Saeedi, Morteza Saheb Zamani, Mehdi Sedighi, Zahra Sasanian \\
\normalsize Quantum Design Automation Lab\\
\normalsize Department of Computer Engineering and Information Technology\\
\normalsize Amirkabir University of Technology\\
\normalsize Tehran, Iran \\
\normalsize \texttt{\{msaeedi, szamani, msedighi, sasanian\}@aut.ac.ir}
 }
 \date{}
\maketitle
\begin{abstract}
Reversible logic has applications in various research areas including signal processing, cryptography and quantum computation. In this paper, direct NCT-based synthesis of a given $k$-cycle in a cycle-based synthesis scenario is examined. To this end, a set of seven building blocks is proposed that reveals the potential of direct synthesis of a given permutation to reduce both quantum cost and average runtime. To synthesize a given large cycle, we propose a decomposition algorithm to extract the suggested building blocks from the input specification. Then, a synthesis method is introduced which uses the building blocks and the decomposition algorithm. Finally, a hybrid synthesis framework is suggested which uses the proposed cycle-based synthesis method in conjunction with one of the recent NCT-based synthesis approaches which is based on Reed-Muller (RM) spectra.\\
The time complexity and the effectiveness of the proposed synthesis approach are analyzed in detail. Our analyses show that the proposed hybrid framework leads to a better quantum cost in the worst-case scenario compared to the previously presented methods. The proposed framework always converges and typically synthesizes a given specification very fast compared to the available synthesis algorithms. Besides, the quantum costs of benchmark functions are improved about 20\% on average (55\% in the best case).
\end{abstract}

\newtheorem{theorem}{Theorem}[section]
\newtheorem{lemma}{Lemma}[section]
\newtheorem{example}{Example}[section]
\newproof{pf}{Proof}

\section{Introduction}

Reversible computing deals with any computational process that is time-invertible, meaning that the process can also be computed backward through time. A necessary condition for reversibility is that the transition function applied to map inputs onto outputs works as a one-to-one function to have a unique output assignment for each input pattern. Generally, conventional logic gates other than NOT are not reversible, as their inputs cannot be determined from the related outputs uniquely.

One of the motivations for research on reversible computing is that it offers a potential way to improve the energy efficiency of computers beyond the fundamental Landauer limit introduced in 1961 \cite{Landauer61}. Landauer proved that using conventional irreversible logic gates leads to at least $kT \times ln2$ energy dissipation per irreversible bit operation, regardless of the underlying circuit, where $k$ is Boltzmann's constant, and $T$ is the temperature of the environment. In 1973, Bennett stated that to avoid power dissipation in a circuit, the circuit must be built from reversible gates \cite{Bennett73}. This has made reversible computing an attractive option for low-power design \cite{Zhirnov03}, \cite{Schrom98}. Additionally, the field of reversible computing has received considerable attention in quantum computing as each quantum gate is reversible in nature \cite{Nielsen00}.

Among various open research problems related to the field of reversible computing, reversible logic synthesis, defined as the ability to generate an efficient circuit from a given arbitrary-size specification, is considered as a stepping-stone towards realization of useful reversible hardware. As a result, working on synthesis methods for reversible circuits has received a significant attention recently (for examples see \cite{MaslovTODAES07}, \cite{MaslovTCAD08} and \cite{Gupta06}). As loop and fanout are not allowed in reversible circuits, and each gate must have the same number of inputs and outputs with unique input/output assignments in the transition function, mature irreversible synthesis algorithms cannot be directly applied to reversible circuits.

To synthesize a given reversible specification, the authors of \cite{Shende03} proposed a synthesis algorithm based on NOT, CNOT and Toffoli gates which represents a given permutation as a product of pairs of disjoint transpositions (2-cycles) and synthesizes each pair subsequently. A general permutation should be decomposed into a set of 2-cycles to be synthesizable using their approach. In this paper, a $k$-cycle-based synthesis method is proposed and analyzed in detail. We show that direct synthesis of large cycles in a cycle-based synthesis scenario can lead to a significant reduction in quantum cost. In order to achieve this, several building blocks (BBs) and synthesis algorithms are proposed to be used in the proposed $k$-cycle-based synthesis method. In addition, a decomposition algorithm for the synthesis of a general large cycle considering the suggested building blocks is introduced and analyzed. Based on the characterization of the proposed synthesis method, a hybrid synthesis framework, which uses the cycle-based synthesis approach in conjunction with one of the recent methods \cite{MaslovTODAES07}, is also presented. Furthermore, the average-case and worst-case quantum costs of the proposed synthesis framework are experimented and analyzed in detail.

The main contributions of this paper are as follows.
\begin{itemize}
\item The analysis of cycle-based synthesis approach and its usefulness in synthesizing reversible functions with different characterizations,
\item A $k$-cycle-based synthesis method with guaranteed convergence,
\item A hybrid synthesis framework based on the proposed $k$-cycle-based synthesis method together with the method of \cite{MaslovTODAES07},
\item The improved quantum cost in the worst-case scenario compared to the previously presented methods,
\item Better average quantum costs for available benchmark functions in the NCT library,
\item Improved average runtime compared to the present synthesis algorithms with favorable synthesis costs.
\end{itemize}

The rest of this paper is organized as follows: In Section \ref{sec:basic_concepts}, basic concepts are introduced. The proposed cycle-based synthesis method is presented in Section \ref{sec:our_method} where the building blocks and their synthesis algorithms are proposed in Subsection \ref{sec:BB}, the decomposition algorithm and the $k$-cycle-based synthesis method are explained in Subsection \ref{sec:decomposition}, and the worst-case analysis of the proposed cycle-based approach is discussed in Subsection \ref{sec:worst_case}. Experimental results and the hybrid synthesis framework are proposed in Section \ref{sec:exper} and finally, Section \ref{sec:conclusion} concludes the paper.

\section{Preliminaries} \label{sec:basic_concepts}
Let $A$ be a set and define $f: A \rightarrow A$ as a one-to-one and onto transition function. The function $f$ is called a \emph{permutation function} as applying $f$ to $A$ leads to a set with the same elements of $A$ and probably in a different order. If $A={1, 2, 3, \cdots, m}$ there exist two elements $a_i$ and $a_j$ belonging to $A$ such that $f(a_i)=a_j$. In addition, a \emph{$k$-cycle} with \emph{length} $k$ is denoted as $(a_1, a_2, \cdots, a_k)$ which means that $f(a_1)=a_2$, $f(a_2)=a_3$, ..., and $f(a_k)=a_1$. A given $k$-cycle $(a_1, a_2, \cdots, a_k)$ could be written in many different ways such as $(a_2, a_3, \cdots, a_k, a_1)$. A cycle with length 2 is called \emph{transposition}.

Cycles $c_1$ and $c_2$ are called \emph{disjoint} if they have no common members, i.e., $\forall a_i \in c_1, a_i \notin c_2$ and vice versa. Any permutation can be written uniquely, except for the order, as a product of disjoint cycles. If two cycles $c_1$ and $c_2$ are disjoint, they can \emph{commute}, i.e., $c_1 c_2= c_2 c_1$. In addition, a cycle may be written in different ways; as a product of transpositions and using different numbers of transpositions. A cycle (or a permutation) is called \emph{even} if it can be written as an even number of transpositions. A similar definition is introduced for an \emph{odd} cycle. Although there may be too many ways to decompose a given cycle into a set of transpositions, the parity of the number of transpositions used stays the same, i.e., all resulted decompositions have the same even/odd number of transpositions. A $k$-cycle is odd (even) if $k$ is even (odd).

An $n$-input, $n$-output, fully specified Boolean function is \emph{reversible} if it maps each input pattern to a unique output pattern. In this paper, $n$ is particularly used to refer to the number of inputs/outputs in a circuit. A \emph{gate} is called reversible if it realizes a reversible function. A \emph{generalized Toffoli gate} C$^m$NOT $(x_1, x_2, \cdots, x_{m+1})$ passes the first $m$ lines unchanged. These lines are referred to \emph{control lines}. This gate flips the $(m+1)^{th}$ line if and only if the control lines are all one. Therefore, the generalized Toffoli gate works as follows: $x_{i(out)}=x_i (i<m+1), x_{m+1(out)}=x_1 x_2 \cdots x_m \oplus x_{m+1}$. For $m=0$ and $m=1$, the gates are called \emph{NOT}(\emph{N}) and \emph{CNOT}(\emph{C}), respectively. For $m=2$, the gate is called C$^2$NOT or \emph{Toffoli}(\emph{T}). These three gates compose the universal NCT library and are used in quantum computation frequently \cite{Nielsen00}. Outputs that are not required in the function specification are considered as \emph{garbage} or \emph{auxiliary} bits. The number of elementary gates required for simulating a given gate is called \textit{quantum cost}.

\begin{figure}[!tb]
	\centering
		\includegraphics[height=4cm]{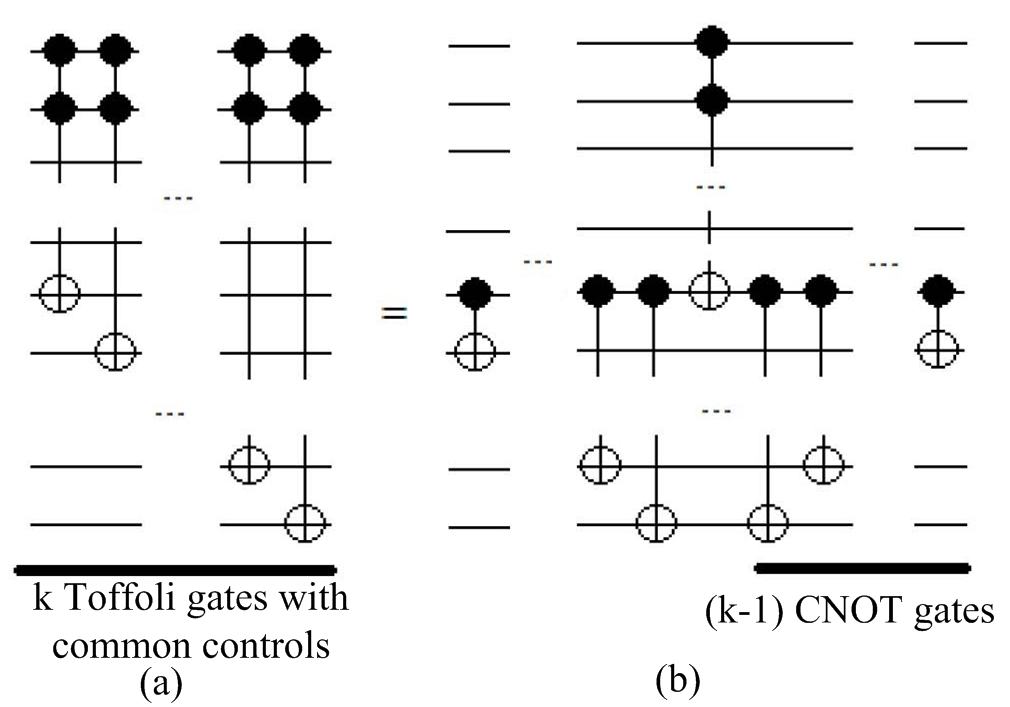}
		\caption{Construction of $k$ Toffoli gates with common controls}
	\label{Fig:1}
\end{figure}

It has been shown that for $n \geq 5$ and $m \in \{3,4, \cdots, \left\lfloor n/2 \right\rceil\}$, a C$^m$NOT gate can be simulated by a linear-size circuit which contains $12m-22$ elementary gates. In addition, for $n \geq 7$, a C$^{n-2}$NOT gate can be simulated by $24n-88$ elementary gates with no auxiliary bits \cite{MaslovTCAD08}. On the other hand, a C$^{n-1}$NOT gate can be simulated with an exponential cost $2^n-3$ if no garbage line is available \cite{Barenco95}. To avoid the exponential size and the need for a large number of elementary gates, several researchers used an extra garbage line for an efficient simulation of C$^{n-1}$NOT gate (e.g., \cite{MaslovTODAES07}). Generally, the number of available bits is very restricted in today's reversible and quantum implementations \cite{Negrevergne06}. Therefore, for two circuits with equal linear costs, the one without garbage line is preferred. The implementation of $k$ Toffoli gates with common controls can be done by $2k+3$ elementary gates as illustrated in Fig. \ref{Fig:1} \cite{MaslovTCAD05}. Note that a Toffoli gate has the cost of 5 whereas NOT and CNOT gates have unit costs.

The authors of \cite{Shende03} proposed an NCT-based synthesis method which applies NOT, Toffoli, CNOT and Toffoli gates in order (the $T|C|T|N$ synthesis method) to synthesize a given permutation. For the last Toffoli part, the authors proposed a synthesis algorithm that maps distinct $a$, $b$, $c$ and $d$ ($a, b, c, d  \neq 0, 2^i$ \label{Reviwer:21} to have at least two ones in their binary representations) to $2^n-4$, $2^n-3$, $2^n-2$ and $2^n-1$ using a circuit called $\pi$ by at most $5n-2$ Toffoli gates. Then, the permutation $(2^n-4, 2^n-3)$ $(2^n-2, 2^n-1)$ is implemented by a circuit, $\kappa_0$, using $8(n-5)$ Toffoli gates and finally, the reversed $\pi$ circuit, i.e.,  $\pi^{-1}$, is applied to transform $2^n-4$, $2^n-3$, $2^n-2$ and $2^n-1$ into $a$, $b$, $c$ and $d$, respectively. Therefore, the $\pi \kappa_0 \pi^{-1}$ circuit implements the permutation $(a, b) (c, d)$ where $a, b, c, d \neq 0, 2^i$ by at most $18n-44$ Toffoli gates.

In contrast, a given $k$-cycle $f$=$(x_0, x_1, x_2, \cdots, x_k)$ is decomposed into a set of transpositions in \cite{Shende03} by using the decomposition pattern $f$=$(x_0, x_1)$ $(x_{k-1}, x_k)$ $(x_0, x_2, x_3, \cdots, x_{k-1})$, recursively. Subsequently, each pair of the transpositions is implemented using the $\pi \kappa_0 \pi^{-1}$ circuit. The proposed approach leads to at most $n$ NOT gates, $n^2$ CNOT gates and $3(2^n+n+1)(3n-7)$ Toffoli gates \cite{Shende03}. An extension of \cite{Shende03} was suggested in \cite{Prasad06} which produced better quantum cost by applying the unit-cost NOT and CNOT gates instead of using Toffoli gates with cost 5 in many situations.

In this paper, the $\pi \kappa_0 \pi^{-1}$ circuit is improved by a $k$-cycle-based synthesis method.
For the rest of this paper, we use the same notations as \cite{Shende03} for the $\kappa_0$, $\pi$, and $\pi^{-1}$ circuits. In all figures, the $(n-1)^{th}$ bit represents the most significant bit (MSB) and is shown as the top line in the circuit representations. Similarly, the $0^{th}$ bit represents the least significant bit (LSB) and is shown as the bottom line in the circuit representations.

\section{$k$-Cycle-Based Synthesis Method} \label{sec:our_method}
\subsection{Building Blocks} \label{sec:BB}
In this subsection, direct synthesis algorithms for seven suggested building blocks (i.e., a pair of 2-cycles, a single 3-cycle, a pair of 3-cycles, a single 5-cycle, a pair of 5-cycles, a single 2-cycle (4-cycle) followed by a single 4-cycle (2-cycle), and a pair of 4-cycles) are introduced and evaluated.
Consider a given 5-cycle $f$=$(a_1, a_2, a_3, a_4, a_5)$ defined in a 7-bit circuit. Assume that $a_1$, $a_2$, $a_3$, $a_4$, and $a_5$ are neither 0 nor $2^i$ to have at least two ones in their binary representations. Applying the decomposition method of \cite{Shende03} leads to $(a_1, a_2)$ $(a_3, a_4)$ $(a_1, a_3)$ $(a_1, a_5)$ transpositions which could be implemented by at most $3 \times (18n-44)=54n-132$ Toffoli gates with cost $270n-660$. However, we will show that a direct 5-cycle implementation of $f$ reduces the total quantum cost to at most $60n-144$.

The proposed synthesis method treats the zero and $2^i$ terms different from the remaining terms. The first group is handled in a pre-process stage similar to the method presented in \cite{Shende03}. For an arbitrary $k$-cycle $(a_1, a_2, \cdots, a_k)$ in the second group, it can be assumed that $a_1, a_2, \cdots a_k \neq 0, 2^i$ and $a_1$ $\neq$ $a_2$ $\neq$ $\cdots$ $\neq$ $a_k$. Throughout this paper, the binary representation is used where CNOT and Toffoli control bits are demonstrated in bold face and the rightmost bit is numbered as the $0^{th}$ (least significant) bit. In order to use the decomposition algorithm proposed in \cite{MaslovTCAD08}, we assume that $n\geq7$.

\begin{figure*}[t]
\begin{minipage}[t]{.5\textwidth}
\centering
\includegraphics[height=3.2cm]{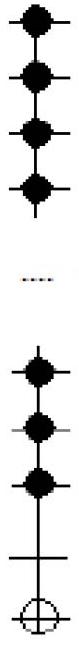}
\caption{The $\kappa_{0(2,2)}$ circuit}\label{Fig:2}
\end{minipage}
\begin{minipage}[t]{.5\textwidth}
\centering
\includegraphics[height=2.7cm]{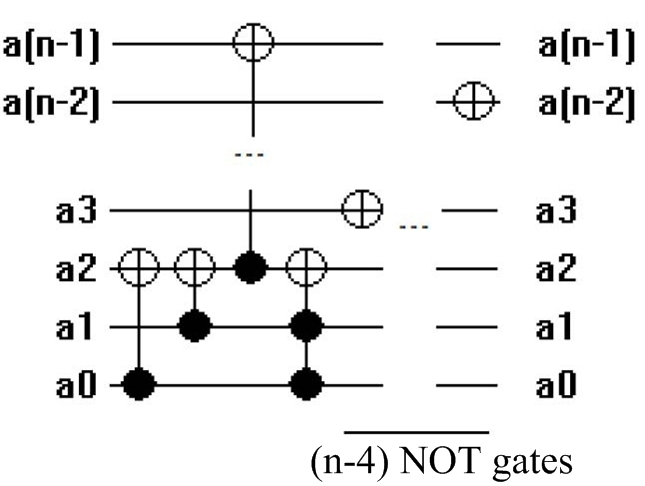}
\caption{The circuit of Theorem \ref{theorem:1}}\label{Fig:3}
\end{minipage}
\end{figure*}

\begin{lemma} \label{lemma:1}
The $\kappa_{0(2,2)}$ circuit $($Fig. \ref{Fig:2}$)$ creates a pair of 2-cycles $(2^n-4, 2^n-3)$ $(2^n-2, 2^n-1)$ by $24n-88$ elementary gates.
\end{lemma}

\begin{pf}
Lemma 20 of \cite{Shende03} proves the correspondence between the $\kappa_{0(2,2)}$ circuit and above cycles. As for the cost, it can be obtained by applying the results of \cite{MaslovDATE05}.
\end{pf}

\begin{figure}[tb]
	\centering
\includegraphics[height=3.5cm]{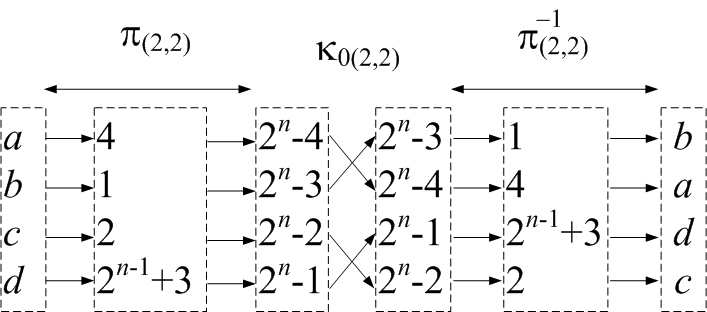}
\caption{Synthesis of an arbitrary pair of 2-cycles $(a, b)$ $(c, d)$}\label{Fig:3-1}
\end{figure}

According to Lemma \ref{lemma:1}, the $\kappa_{0(2,2)}$ circuit implements the particular pair of 2-cycles $(2^n-4, 2^n-3)$ $(2^n-2, 2^n-1)$. In order to implement an arbitrary pair $(a, b)$ $(c, d)$, the circuit is divided into five parts as follows. First, the terms $a$, $b$, $c$ and $d$ are changed to 4, 1, 2 and $2^{n-1}+ 3$, respectively. Note that the first three terms have only one 1 in their binary representations. As shown in the following theorem, this characterization is used during the synthesis of a pair of 2-cycles. Second, a circuit is applied to change 4, 1, 2 and $2^{n-1}+ 3$ to $2^n-4$, $2^n-3$, $2^n-2$ and $2^n-1$ (i.e., the terms used in $\kappa_{0(2,2)}$ circuit), correspondingly. Afterward, the $\kappa_{0(2,2)}$ circuit is used which changes $2^n-4$, $2^n-3$, $2^n-2$ and $2^n-1$ to $2^n-3$, $2^n-4$, $2^n-1$ and $2^n-2$, respectively. Applying the second and the first sub-circuits in the reverse order puts unwanted terms (i.e., all terms except $a$, $b$, $c$ and $d$) back to their original locations and implements the given pair of 2-cycles $(a, b)$ $(c, d)$. Fig. \ref{Fig:3-1} demonstrates the complete synthesis scenario. Theorem \ref{theorem:1} discusses the synthesis of an arbitrary pair of 2-cycles in more details. The synthesis procedures for other cycles are similar to the one explained here as shown later.

\begin {theorem}  \label {theorem:1}
$($Syn$_{2,2}$ method$)$: An arbitrary pair of 2-cycles $(a, b)$ $(c, d)$ can be simulated by at most $34n-64$ elementary gates.
\end {theorem}

\begin{pf}
Since $a$, $b$, $c$ and $d$ are neither 0 nor $2^i$, they should have at least two ones in their binary representations. Assume that the ${c_1}^{th}$ bit of $a$ is 1. One can use at most one CNOT gate whose control is on $c_1$ to set the $2^{nd}$ bit of $a$ to 1. Subsequently, by using at most $n-1$ CNOT gates whose controls are on the $2^{nd}$ bit other bits can be set to 0 for converting $a$ to 4 (i.e., $0 \cdots 0\textbf{1}00$). Assume that after applying these gates, $b$, $c$ and $d$ are changed to $b'$, $c'$, and $d'$, respectively. Since $b'$ should have at least one 1 namely at position $c_2$ $(c_2 \neq 2)$, $b'$ can be converted to 1 (i.e., $0 \cdots 0\textbf{1}$) by at most $n$ CNOT gates using a similar approach. Then, $c'$ and $d'$ may be changed to new numbers $c''$ and $d''$, respectively without changing 4.

Subsequently, $c''$ can be converted to 2 (i.e., $0 \cdots 0\textbf{1}0$) by at most one Toffoli gate and $n-1$ CNOT gates with no effects on 4 and 1. Finally, the last term can be converted to $2^{n-1}+3$ (i.e., $\textbf{1}0 \cdots 011$) by at most one Toffoli gate and $n-1$ CNOT gates with no effect on the previous terms again. Therefore, at most $4n+8$ elementary gates are required to transform $a$, $b$, $c$ and $d$ into 4, 1, 2 and $2^{n-1}+ 3$, respectively. Now, the circuit shown in Fig. \ref{Fig:3} should be applied to change 4, 1, 2 and $2^{n-1}+ 3$ to the terms used in $\kappa_{0(2,2)}$ circuit (Lemma \ref{lemma:1}). Considering the applied gates (at most $5n + 12$ elementary gates), the terms $a$, $b$, $c$ and $d$ are changed to $2^n-4$, $2^n-3$, and $2^n-2$ and $2^n-1$, respectively (i.e., the $\pi_{2,2}$ circuit). Now, by using the $\kappa_{0(2,2)}$ circuit with the cost of $24n-88$, the pair of 2-cycles $(2^n-4, 2^n-3)$ $(2^n-2, 2^n-1)$ is implemented. Applying the $\pi_{2,2}^{-1}$ circuit changes $2^n-4$, $2^n-3$, and $2^n-2$ and $2^n-1$ to $a$, $b$, $c$, and $d$, respectively. In addition, the circuit $\pi_{2,2}^{-1}$ puts other unwanted terms back to their original locations. Therefore, by at most $34n-64$ elementary gates, the pair of 2-cycles $(a, b)$ $(c, d)$ can be implemented.	
\end{pf}

\begin{figure}[!tb]
	\centering
\includegraphics[height=3.8cm]{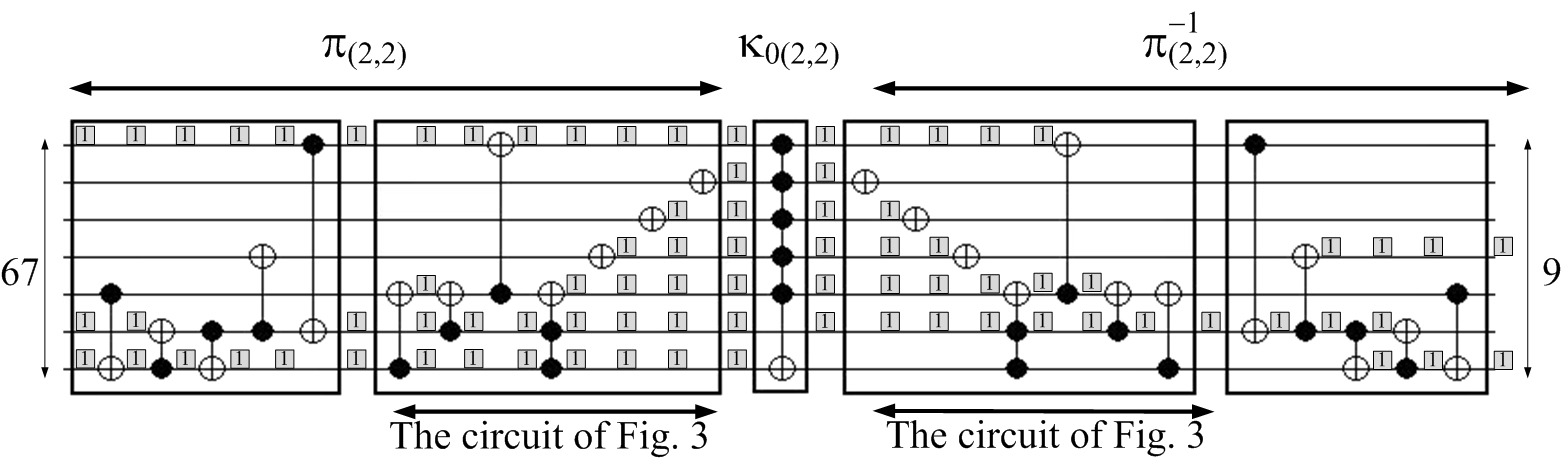}
\caption{The circuit of Example \ref{example:1}}\label{Fig:4}
\end{figure}

\begin{example} \label{example:1}
Assume that the pair of 2-cycles $(5, 3)$ $(9, 67)$ should be implemented in a circuit over 7 bits (i.e., n=7). According to the proof of Theorem \ref{theorem:1}, the term 5 should be transformed to 4 by a CNOT gate which has no effect on other terms. Similarly, 3 is transformed to 1 by a CNOT gate which changes the term 9 to 11 and 67 to 65. Then, 11 is transformed to 2 by two CNOT gates with no effect on other terms. Finally, 65 is transformed to 67 by a CNOT gate. See the first sub-circuit in Fig. \ref{Fig:4} for more details. Now, the circuit shown in Fig. \ref{Fig:3} should be applied followed by the $\kappa_{0(2,2)}$ circuit. Finally, as illustrated in the last two sub-circuits of Fig. \ref{Fig:4}, the above gates (except $\kappa_{0(2,2)}$ circuit) should be used in the reverse order to construct the complete circuit. In Fig. \ref{Fig:4}, the results of applying all gates on the term 67 are also represented by gray squares where only values 1 are shown for the sake of simplicity. As can be seen, applying all gates changes 67 to 9.
\end{example}

\begin{figure}[t]
\begin{minipage}[t]{.5\textwidth}
\centering
\includegraphics[height=4.2cm]{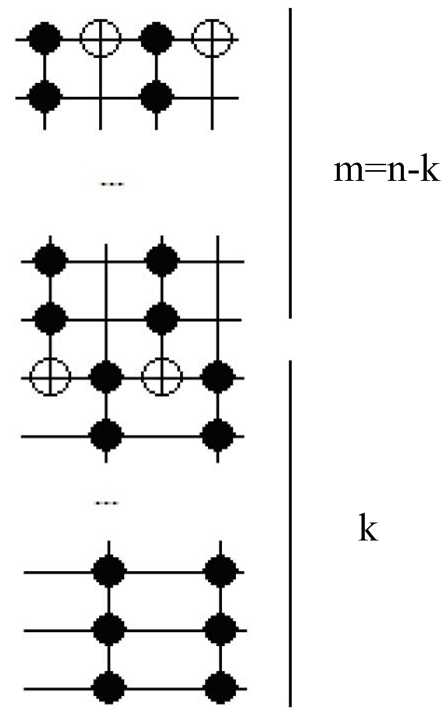}
\caption{The $\kappa_{0(3)}$ circuit}
\label{Fig:5}
\end{minipage}
\begin{minipage}[t]{.5\textwidth}
\centering
\includegraphics[height=3.8cm]{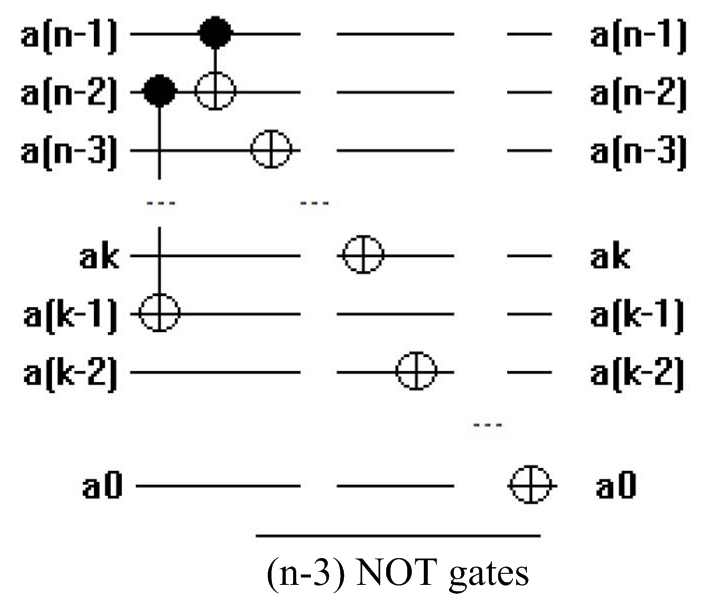}
\caption{The circuit of Theorem \ref{theorem:2}}
\label{Fig:6}
\end{minipage}
\end{figure}

\begin{lemma} \label {lemma:k03}
The $\kappa_{0(3)}$ circuit $($Fig. \ref{Fig:5}$)$ creates the 3-cycle $(2^n-2^{k-1}-1$, $2^n-1$, $2^{n-1}-1)$ by $24n-88$ elementary gates where $k= \left\lceil n/2 \right\rceil$.
\end{lemma}

\begin{pf}
As shown in Fig. \ref{Fig:5}, the gates C$^m$NOT($n-1$, $n-2$, $\cdots$, $k$, $k-1$), C$^k$NOT($0$, $1$, $2$, $\cdots$, $k-1$, $n-1$), C$^m$NOT($n-1$, $n-2$, $\cdots$, $k$, $k-1$), C$^k$NOT($0$, $1$, $2$, $\cdots$, $k-1$, $n-1$) are applied consecutively in the $\kappa_{0(3)}$ circuit. After applying the first C$^m$NOT gate, the locations of $2^k$ minterms (denoted as $\sum_1$=\{$2^n-2^k$, $2^n-2^k+1$, $\cdots$, $2^n-1$\}) are changed. Particularly, $2^n-2^{k-1}-1$ (i.e., $1 \cdots 1\underline{0}1 \cdots 1$ where the underlined 1 is at the $(k-1)^{th}$ position) $\in \sum_1$ is changed to $2^n-1$ (i.e., $1 \cdots 1$)$( \in \sum_1 )$. By applying the C$^k$NOT, the locations of $2^m$ minterms (denoted as $\sum_2$=\{$0 \times 2^k+2^k-1$, $1 \times 2^k+2^k-1$, $\cdots$, $2^{m-1} \times 2^k+2^k-1$=$2^n-1$\} are changed ($2^n-1 \in \sum_1 \cap \sum_2$).
Among them, $2^n-1$ is exchanged with $2^{n-1}-1$  (i.e., $01 \cdots 1$) $\in \sum_2$. Applying the third C$^m$NOT gate puts all $\sum_1$ minterms at their right locations except $2^n-2^{k-1}-1$ and also changes $2^n-1$ to $2^n-2^{k-1}-1$. Finally, the last C$^k$NOT gate corrects the locations of all $\sum_2$ members except $2^{n-1}-1$ and $2^n-1$. Considering all the exchanges, $2^n-2^{k-1}-1$ is changed to $2^n-1$, $2^n-1$ is changed to $2^{n-1}-1$, and $2^{n-1}-1$ is changed to $2^n-2^{k-1}-1$.

For the second part of the lemma, note that the first and the third gates shown in Fig. \ref{Fig:5} can be implemented by $2 \times (12 \times (n- \left\lceil n/2 \right\rceil)-22)$ elementary gates. Similarly, the second and the fourth gates can be implemented by $2 \times (12 \times \left\lceil n/2 \right\rceil-22)$ gates. Therefore, $\kappa_{0(3)}$ is implemented by cost $24n-88$.
\end{pf}

\begin{theorem} \label{theorem:2}
$(Syn_3$ method$)$: An arbitrary 3-cycle $(a, b, c)$ requires at most $32n-82$ elementary gates to be implemented.
\end{theorem}

\begin{pf}
Since $a$, $b$, and $c$ are neither $0$ nor $2^i$, they should have at least two ones in their binary representations. One can use at most $n$ CNOT gates to transform $a$ to $2^{n-1}$ (i.e., $\textbf{1}0 \cdots 0$). After applying these gates, assume that $b$ and $c$ are changed to $b'$ and $c'$, respectively. By using a similar approach, $c'$ can be converted to $2^{n-2}$ (i.e., $0\textbf{1}0 \cdots 0$) by $n$ CNOT gates that may change $b'$ to a new number $b''$ without changing $2^{n-1}$. Finally, converting $b''$ to $2^{n-1} + 2^{k-1}$ (i.e., $10 \cdots 0\textbf{\underline{1}}0 \cdots 0$ where the underlined 1 is at the $(k-1)^{th}$ position) can be done by one Toffoli and $n-1$ CNOT gates with no effects on the previous $2^{n-1}$ and $2^{n-2}$ terms. Therefore, by at most $3n+4$ elementary gates, $a$, $b$ and $c$ are transformed into $2^{n-1}$, $2^{n-1}+2^{k-1}$ and $2^{n-2}$, respectively. Now, the circuit shown in Fig. \ref{Fig:6} should be applied to change the recent terms to the terms used in $\kappa_{0(3)}$ circuit.

Considering the applied gates (at most $4n+3$ elementary gates), the terms $a$, $b$, and $c$ are changed to $2^n-2^{k-1}-1$, $2^n-1$, and $2^{n-1}-1$, respectively (i.e., the $\pi_3$ circuit). By using the $\kappa_{0(3)}$ circuit with cost $24n-88$, the 3-cycle ($2^n-2^{k-1}-1$, $2^n-1$, $2^{n-1}-1$) is implemented. Applying the $\pi_3^{-1}$ circuit changes $2^n-2^{k-1}-1$, $2^n-1$, and $2^{n-1}-1$ to $a$, $b$, and $c$, respectively. Therefore, by at most $32n-82$ elementary gates, the 3-cycle ($a$, $b$, $c$) can be implemented. It is worth noting that a single 3-cycle can be a BB by itself because it is even. As will be shown later, the same is true for a single 5-cycle.
\end{pf}

\begin{figure}[t]
\begin{minipage}[t]{.38\textwidth}
\centering
\includegraphics[height=3.6cm]{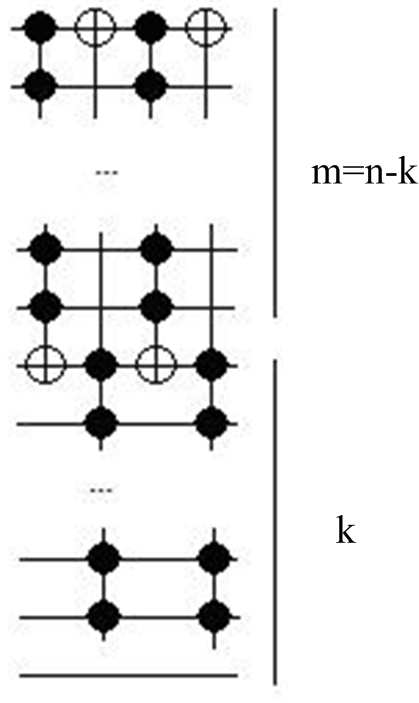}
\caption{The $\kappa_{0(3,3)}$ circuit, $k=\left\lceil n/2 \right\rceil$ }
\label{Fig:7}
\end{minipage}
\begin{minipage}[t]{.62\textwidth}
\centering
\includegraphics[height=4cm]{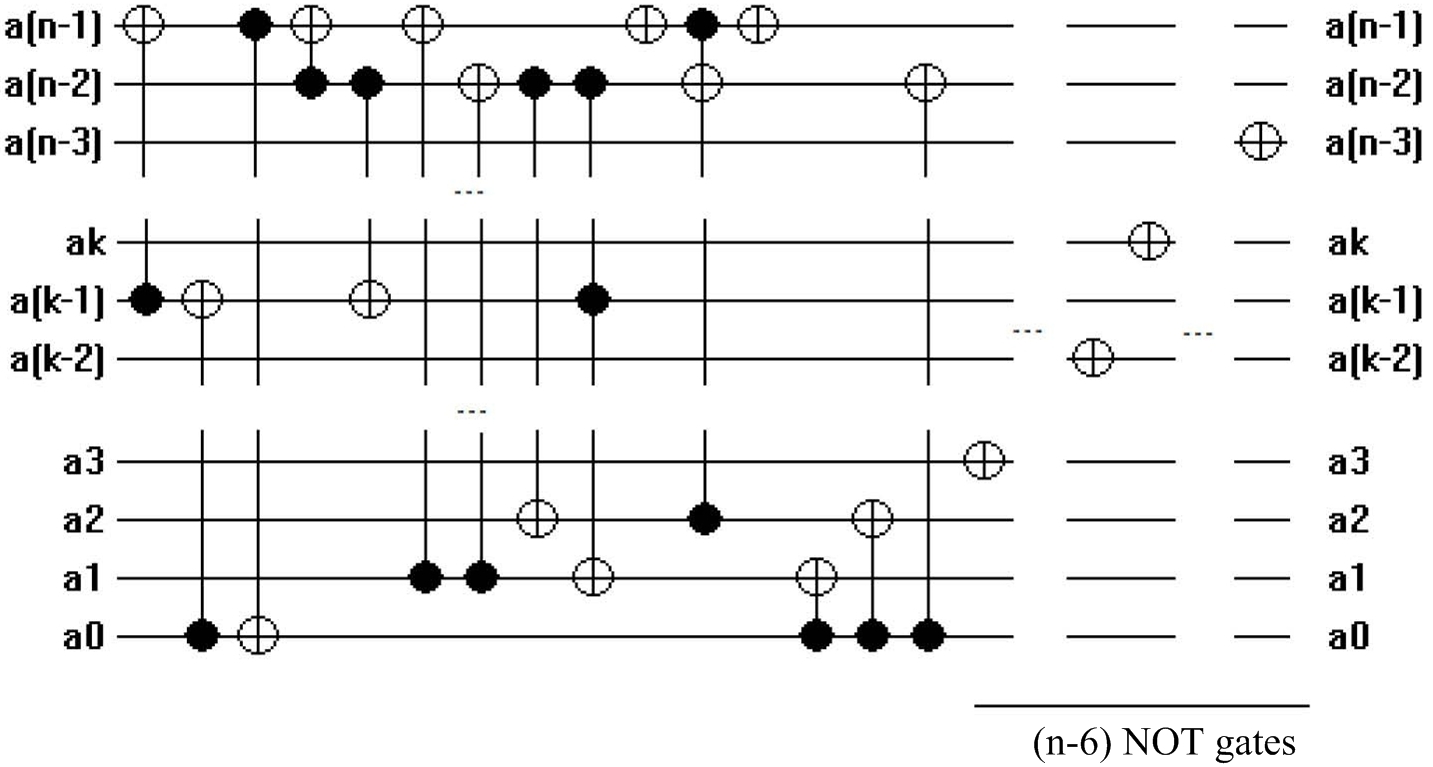}
\caption{The circuit of Theorem \ref{theorem:3} }
\label{Fig:8}
\end{minipage}
\end{figure}

\begin{lemma}
The $\kappa_{0(3,3)}$ circuit $($Fig. \ref{Fig:7}$)$ implements the pair of 3-cycles $(2^n-2^{k-1}-1$, $2^n-1$, $2^{n-1}-1)$ $(2^n-2^{k-1}-2$, $2^n-2$, $2^{n-1}-2)$ by $24n-112$ elementary gates where $k= \left\lceil n/2 \right\rceil $.
\end{lemma}

\begin{pf}
It can be verified that the $\kappa_{0(3,3)}$ circuit differs from the $\kappa_{0(3)}$ circuit in its least significant bit (i.e., the $0^{th}$ bit) which leads to two 3-cycles. The first and the third gates need $12n-44$ elementary gates. The second and the fourth gates need $12n-68$ elementary gates. Therefore, $\kappa_{0(3,3)}$ can be implemented by the cost of $24n-112$ gates.	
\end{pf}

\begin{theorem} \label{theorem:3}
$(Syn_{3,3}$ method$)$: The implementation of an arbitrary pair of 3-cycles $(a$, $b$, $c)$ $(d$, $e$, $f)$ requires at most $38n-46$ elementary gates.
\end{theorem}

\begin{pf}
Use at most $6n+16$ elementary gates to convert $a$ to $2^{n-1}$ (i.e., $\textbf{1}0 \cdots 0$), $b$ to $2^{k-1}$ (i.e., $0 \cdots 0 \textbf{\underline{1}} 0 \cdots 0$ where the underlined 1 is at the $(k-1)^{th}$ position), $c$ to $1$ (i.e., $0 \cdots 0\textbf{1}$), $d$ to $2$ (i.e., $0 \cdots 0\textbf{1}0$), $e$ to $2^{n-2}$ (i.e., $0\textbf{1}0 \cdots 0$), and $f$ to $2^{n-2}+6$ (i.e., $010 \cdots 0\textbf{1}10$), sequentially. Therefore, the terms $a$, $b$, $c$, $d$, $e$, and $f$ are changed to $2^{n-1}$, $2^{k-1}$, $1$, $2$, $2^{n-2}$, and $2^{n-2}+6$, respectively. Note that the terms $a$ and $b$ can be implemented by only CNOT gates. For each of the other terms, at most one Toffoli and $n-1$ CNOT gates should be applied. Now, apply the circuit shown in Fig. \ref{Fig:8}. After applying at most $7n+33$ elementary gates, $a$, $b$, $c$, $d$, $e$, and $f$ are transformed into $2^n-2^{k-1}-2$, $2^n-2$, $2^{n-1}-2$, $2^n-2^{k-1}-1$, $2^n-1$, and $2^{n-1}-1$, respectively (i.e., $\pi_{3,3}$ circuit). By applying $\kappa_{0(3,3)}$ and the reversed $\pi_{3,3}$ circuit, $38n-46$ elementary gates are used and ($a$, $b$, $c$) ($d$, $e$, $f$) is implemented.			
\end{pf}

\begin{figure}[t]
\begin{minipage}[t]{.45\textwidth}
\centering
\includegraphics[height=3.5cm]{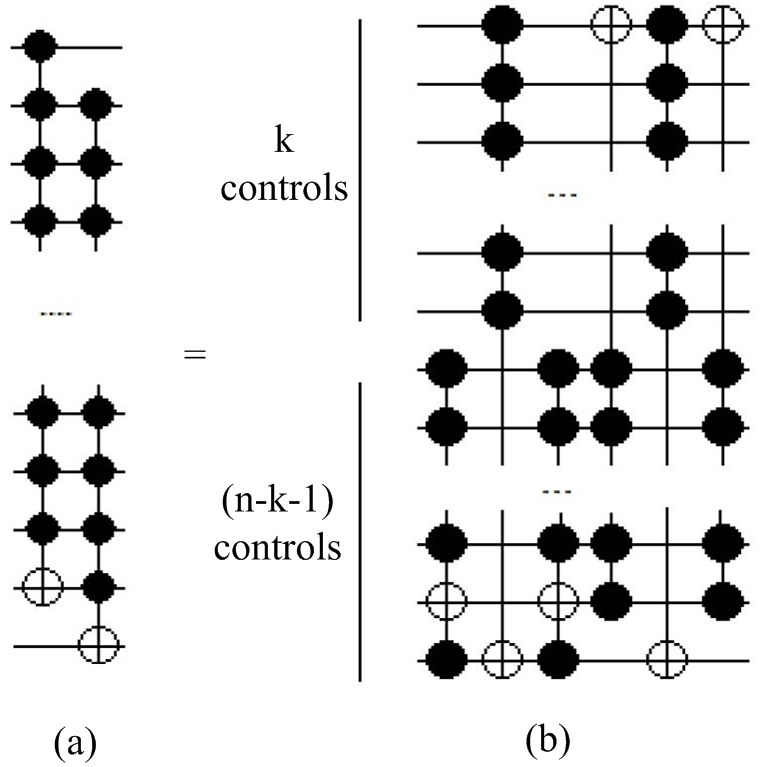}
\caption{The $\kappa_{0(4,2)}$ circuit, $k=\left\lceil n/2 \right\rceil$}
\label{Fig:9}
\end{minipage}
\begin{minipage}[t]{.55\textwidth}
\centering
\includegraphics[height=3.5cm]{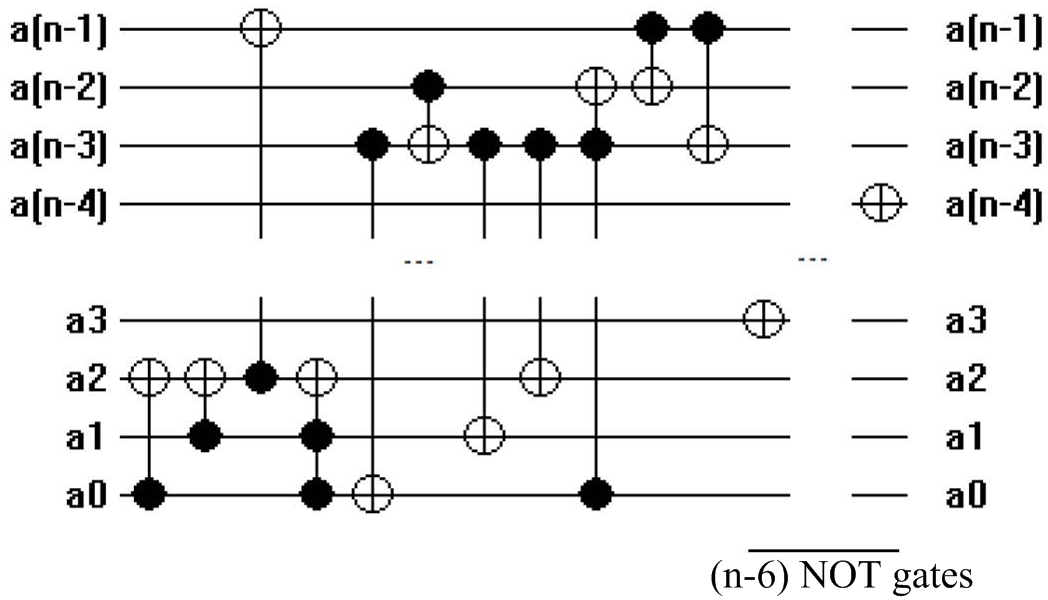}
\caption{The circuit of Theorem \ref{theorem:4}}
\label{Fig:10}
\end{minipage}
\end{figure}

\begin{lemma}
The $\kappa_{0(4,2)}$ circuit $($Fig. \ref{Fig:9}-$a)$ implements the pair $(2^n-4$, $2^n-1$, $2^n-3$, $2^n-2)$ $(2^{n-1}-2$, $2^{n-1}-1)$ by $36n-180$ elementary gates.
\end{lemma}

\begin{pf}
The first C$^{n-2}$NOT($n-1$, $n-2$, $\cdots$, $2$, $1$) gate shown in Fig. \ref{Fig:9}-a changes $2^n-4$, $2^n-3$, $2^n-2$, and $2^n-1$ to $2^n-2$, $2^n-1$, $2^n-4$, and $2^n-3$, respectively. The second C$^{n-2}$NOT($n-2$, $\cdots$ ,$2$, $1$, $0$) changes $2^n-2$, $2^n-1$, $2^{n-1}-2$ and $2^{n-1}-1$ to $2^n-1$, $2^n-2$, $2^{n-1}-1$ and $2^{n-2}-2$, respectively. Considering the gates sequentially leads to the implementation of $\kappa_{0(4,2)}$. The circuit in Fig. \ref{Fig:9}-b can be obtained by applying the Lemma 7.3 of \cite{Barenco95} on each C$^{n-2}$NOT gate of Fig. \ref{Fig:9}-a and canceling the resulted redundant gates. The total number of $36n-180$ elementary gates can be achieved by a summation of the costs of gates in Fig. \ref{Fig:9}-b. 	
\end{pf}

\begin{theorem} \label{theorem:4}
$(Syn_{4,2}$ method$)$: An arbitrary pair $(a$, $b$, $c$, $d)$ $(e$, $f)$ can be implemented by at most $50n-122$ elementary gates.
\end{theorem}

\begin{pf}
Use at most $6n+16$ elementary gates to convert $a$ to 4 (i.e., $0 \cdots 0\textbf{1}00$), $c$ to $1$ (i.e., $0 \cdots 0\textbf{1}$), $d$ to $2$ (i.e., $0 \cdots 0\textbf{1}0$), $e$ to $2^{n-2}$ (i.e., $0\textbf{1}0 \cdots 0$), $f$ to $2^{n-3}$ (i.e., $00\textbf{1}0 \cdots 0$), and $b$ to $2^{n-1}+3$ (i.e., $\textbf{1}0 \cdots 011$), sequentially. Note that the terms $a$ and $c$ can be implemented by only CNOT gates. For each of the other terms, at most one Toffoli and $n-1$ CNOT gates should be applied. Now, apply the circuit shown in Fig. \ref{Fig:10}. After applying at most $7n+29$ elementary gates, the terms $a$, $b$, $c$, $d$, $e$, and $f$ are changed to $2^n-4$, $2^n-1$, $2^n-3$, $2^n-2$, $2^{n-1}-2$, and $2^{n-1}-1$, respectively (the $\pi_{4,2}$ circuit). Then, apply the $\kappa_{0(4,2)}$ and the reversed $\pi_{4,2}$ circuit (i.e., $\pi_{4,2}^{-1}$) to complete the implementation of ($a$, $b$, $c$, $d$) ($e$, $f$) by at most $50n-122$ elementary gates.	
\end{pf}

\begin{figure}[t]
\begin{minipage}[t]{.45\textwidth}
\centering
\includegraphics[height=3.5cm]{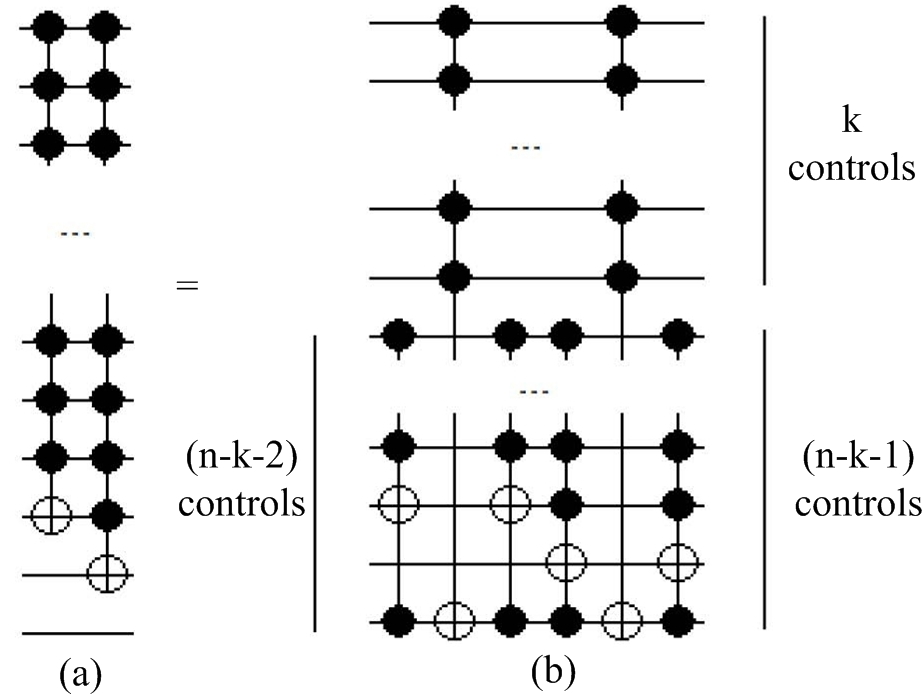}
\caption{The $\kappa_{0(4,4)}$ circuit, $k=\left\lceil n/2 \right\rceil$ }
\label{Fig:11}
\end{minipage}
\begin{minipage}[t]{.55\textwidth}
\centering
\includegraphics[height=3.5cm]{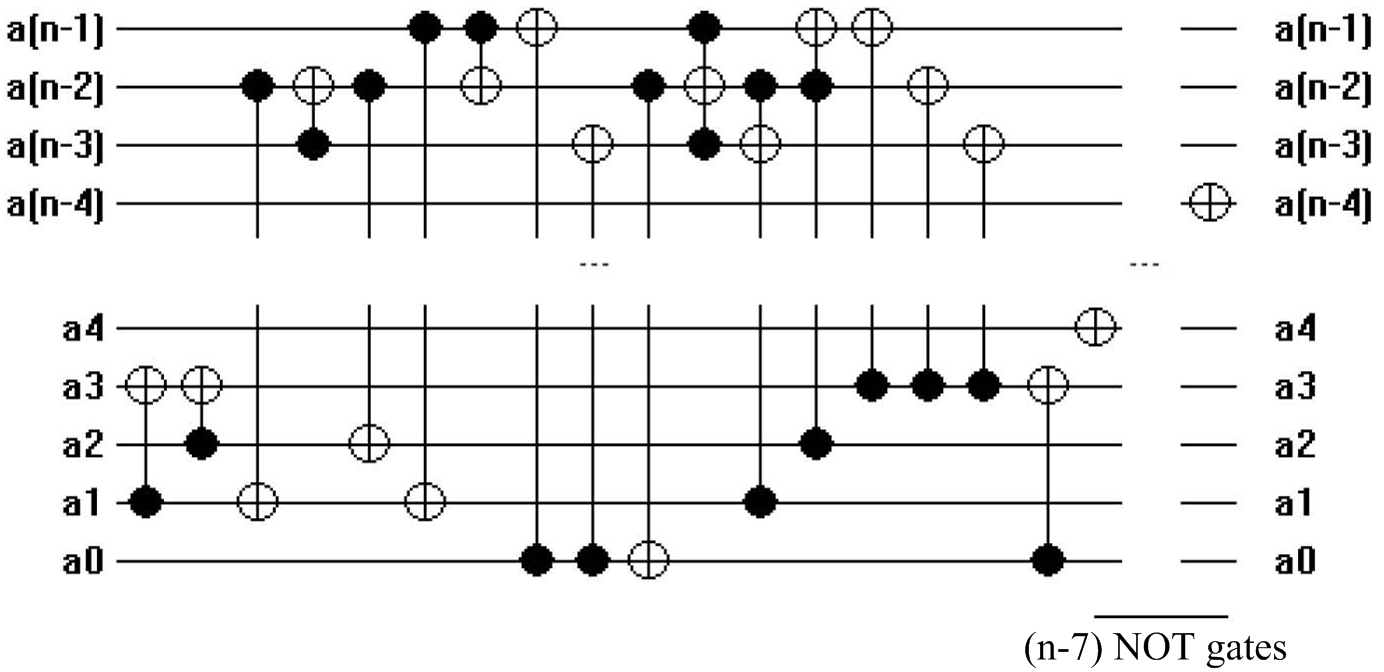}
\caption{The circuit of Theorem \ref{theorem:5}}
\label{Fig:12}
\end{minipage}
\end{figure}

\begin{lemma}
The $\kappa_{0(4,4)}$ circuit $($Fig. \ref{Fig:11}-$a)$ implements $(2^n-8$, $2^n-2$, $2^n-6$, $2^n-4)$ $(2^n-7$, $2^n-1$, $2^n-5$, $2^n-3)$ by cost $36n-228$.
\end{lemma}

\begin{pf}
Consider Fig. \ref{Fig:11}-a. The first C$^{n-3}$NOT($n-1$, $n-2$, $\cdots$, $3$, $2$) gate changes $2^n-8$, $2^n-7$, $2^n-6$ and $2^n-5$ to $2^n-4$, $2^n-3$, $2^n-2$ and $2^n-1$, respectively. The second C$^{n-2}$NOT($n-1$, $n-2$, $\cdots$, $2$, $1$) gate changes $2^n-4$, $2^n-3$, $2^n-2$, and $2^n-1$ to $2^n-2$, $2^n-1$, $2^n-4$, and $2^n-3$, respectively. Considering the gates sequentially leads to the implementation of the cycle. Applying the Lemma 7.3 of \cite{Barenco95} on each gate shown in Fig. \ref{Fig:11}-a and canceling the resulted redundant gates transform Fig. \ref{Fig:11}-a to Fig. \ref{Fig:11}-b. The total number of $36n-228$ elementary gates can be obtained by summation of the costs of gates shown in Fig. \ref{Fig:11}-b.
\end{pf}

\begin{theorem} \label{theorem:5}
$(Syn_{4,4}$ method$)$: An arbitrary pair $(a$, $b$, $c$, $d)$ $(e$, $f$, $g$, $h)$ can be implemented by at most $56n-126$ elementary gates.
\end {theorem}

\begin{pf}
Use at most $9n+22$ elementary gates to sequentially convert $a$ to $8$ (i.e., $0 \cdots 0\textbf{1}000$), $c$ to $2$ (i.e., $0 \cdots 0\textbf{1}0$), $d$ to $4$ (i.e., $0 \cdots 0\textbf{1}00$), $e$ to $1$ (i.e., $0 \cdots 0\textbf{1}$), $f$ to $2^{n-2}$ (i.e., $0\textbf{1}0 \cdots 0$), $g$ to $2^{n-1}$ (i.e., $\textbf{1}0 \cdots 0$), $h$ to $2^{n-3}$ (i.e., $00\textbf{1}0 \cdots 0$) and $b$ to $14$ (i.e., $0 \cdots 0\textbf{11}10$). Note that $a$ and $c$ can be transformed to $8$ and $2$ by only CNOT gates, respectively. In addition, for each term $d$, $e$, $f$, $g$, and $h$ at most one Toffoli and $n-1$ CNOT gates should be used. For the last term $b$ at most two Toffoli gates should be used to set the $2^{nd}$ and $3^{rd}$ bits to $1$. Then, at most $n-2$ Toffoli gates should be applied to set the $1^{st}$ bit to $1$ and the $i^{th}$ bit to $0$ where $0 \leq i \leq n-1$, $i \neq 1, 2, 3$. The $n-2$ Toffoli gates can be implemented by cost $2(n-2)+3$ (see Fig. \ref{Fig:1}) since all Toffoli gates use the same control lines (i.e., the $2^{nd}$ and $3^{rd}$ bits). Note that for $n \geq 8$, the term $b$ can also be implemented by at most one Toffoli and $n-1$ CNOT gates. Now, apply the circuit shown in Fig. \ref{Fig:12}. After applying at most $10n+51$ elementary gates, $a$, $b$, $c$, $d$, $e$, $f$, $g$, and $h$ are changed to $2^n-8$, $2^n-2$, $2^n-6$, $2^n-4$, $2^n-7$, $2^n-1$, $2^n-5$, and $2^n-3$, respectively ($\pi_{4,4}$). Then, apply $\kappa_{0(4,4)}$ and  $\pi_{4,4}^{-1}$ to complete the implementation of ($a$, $b$, $c$, $d$) ($e$, $f$, $g$, $h$) by at most $56n-126$ elementary gates.	
\end{pf}

\begin{figure}[t]
\begin{minipage}[t]{.5\textwidth}
\centering
\includegraphics[height=3.4cm]{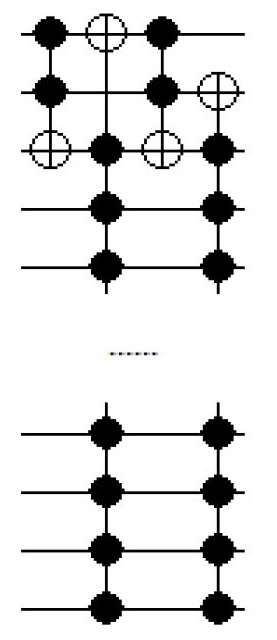}
\caption{The $\kappa_{0(5)}$ circuit}
\label{Fig:13}
\end{minipage}
\begin{minipage}[t]{.5\textwidth}
\centering
\includegraphics[height=2.8cm]{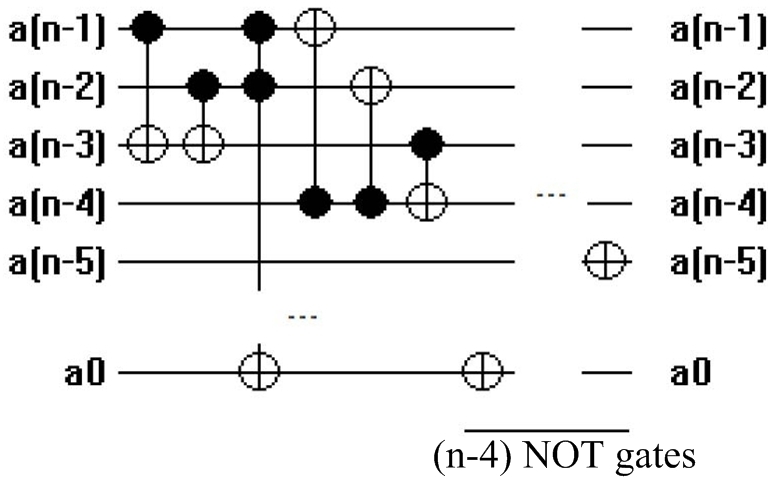}
\caption{The circuit of Theorem \ref{theorem:6}}
\label{Fig:14}
\end{minipage}
\end{figure}

\begin{lemma}
The $\kappa_{0(5)}$ circuit $($Fig. \ref{Fig:13}$)$ implements the 5-cycle $(2^{n-2}-1$, $2^n-1$, $2^n-2^{n-2}-1$, $2^{n-1}-1$, $2^n-2^{n-3}-1)$ with cost $48n-166$.
\end{lemma}

\begin{pf}
As illustrated in Fig. \ref{Fig:13}, four gates T($n-1$, $n-2$, $n-3$), C$^{n-2}$NOT($0$, $\cdots$, $n-3$, $n-1$), T($n-1$, $n-2$, $n-3$), C$^{n-2}$NOT($0$, $\cdots$, $n-1$, $n-2$) are applied sequentially. After applying the first Toffoli gate, the locations of $2^{n-2}$ minterms (i.e., $\sum_1$ = \{$2^n-2^{n-2}$, $2^n-2^{n-2}+1$, $\cdots$, $2^n-1$\}) are changed. Mainly, $2^n-2^{n-3}-1$ (i.e., $1101 \dots 1$) $\in \sum_1$ is changed to $2^n-1$ ($\in \sum_1$). After the second C$^{n-2}$NOT, the locations of $4$ minterms (denoted as $\sum_2$=\{$2^{n-2}-1$, $2^{n-1}-1$, $2^n-2^{n-2}-1$, $2^n-1$\}) are changed (where $2^n-1 \in \sum_1 \cap \sum_2$). Among them, $2^n-1$ is changed to $2^{n-1}-1 \in \sum_2$, and $2^{n-1}-1$ is changed to $2^n-1$. Applying the third Toffoli gate puts all $\sum_1$ minterms at their right locations except $2^n-2^{n-3}-1$. In addition, it changes $2^n-1$ to $2^n-2^{n-3}-1$. Finally, the last C$^{n-2}$NOT gate changes the locations of four minterms as $2^{n-1}-1$ to $2^{n-2}-1$, $2^n-1$ to $2^n-2^{n-2}-1$, $2^n-2^{n-2}-1$ to $2^n-1$, and $2^{n-2}-1$ to $2^{n-1}-1$. Considering all minterm exchanges, it can be verified that the 5-cycle $\kappa_{0(5)}$ is implemented by the circuit of Fig. \ref{Fig:13}. The total number of $48n-166$ elementary gates can be obtained by a summation of the costs of gates in Fig. \ref{Fig:13}.	
\end{pf}

\begin{theorem} \label{theorem:6}
$(Syn_5$ method$)$: An arbitrary 5-cycle $(a$, $b$, $c$, $d$, $e)$ can be implemented by at most $60n-130$ elementary gates.
\end {theorem}

\begin{pf}
Use at most $5n+12$ elementary gates to convert $a$ to $2^{n-3}$ (i.e., $00\textbf{1}0 \cdots 0$), $d$ to $2^{n-2}$ (i.e., $0\textbf{1}0 \cdots 0$), $c$ to $2^{n-1}$ (i.e., $\textbf{1}0 \cdots 0$), $e$ to $2^{n-4}$ (i.e., $000\textbf{1}0 \cdots 0$) and $b$ to $2^{n-1}+2^{n-2}+2^{n-3}+1$ (i.e., $1110 \cdots 0\textbf{1}$), sequentially. Note that $a$ and $d$ can be transformed to $2^{n-3}$ and $2^{n-2}$ by only CNOT gates, respectively. For each of the other terms at most one Toffoli and $n-1$ CNOT gates should be used. Then, apply the circuit shown in Fig. \ref{Fig:14}. After using the applied gates (at most $6n+18$ elementary gates), the terms $a$, $b$, $c$, $d$, and $e$ are changed to $2^{n-2}-1$, $2^n-1$, $2^n-2^{n-2}-1$, $2^{n-1}-1$, and $2^n-2^{n-3}-1$, respectively ($\pi_5$). Therefore, by applying the $\kappa_{0(5)}$ circuit and the $\pi_5^{-1}$ circuit, the 5-cycle ($a$, $b$, $c$, $d$, $e$) is implemented by at most $60n-130$ elementary gates.
\end{pf}

\begin{figure}[t]
\centering
\includegraphics[height=4cm]{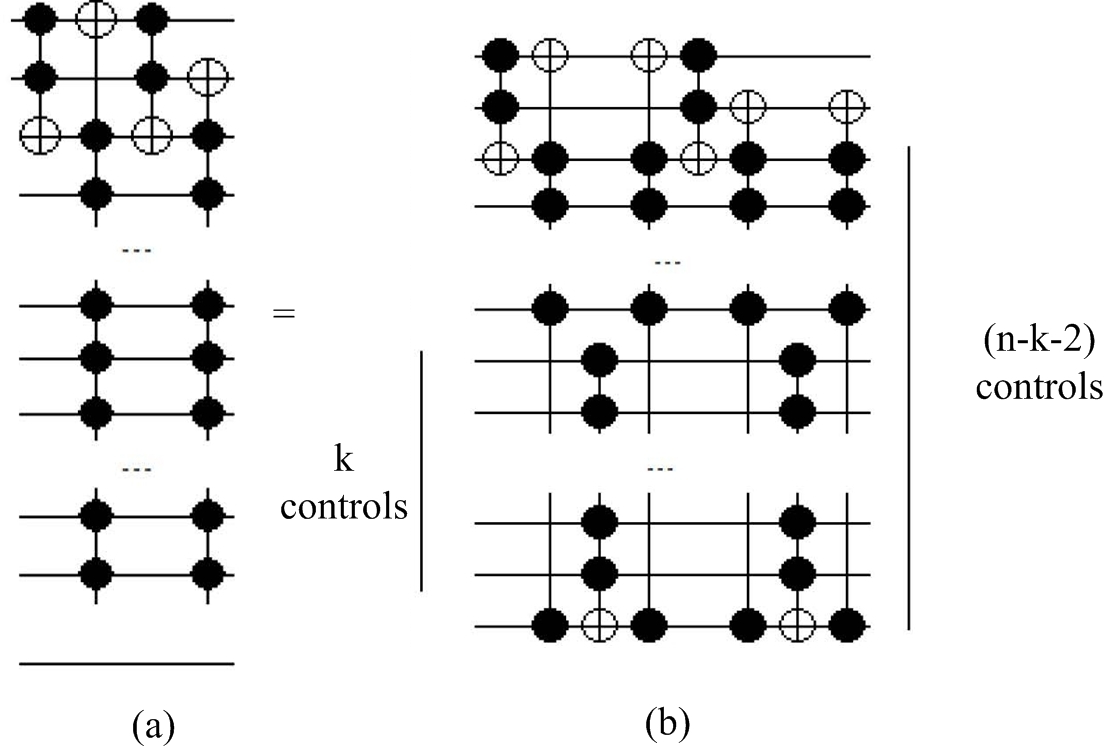}
\caption{The $\kappa_{0(5,5)}$ circuit, $k=\left\lceil (n-1)/2 \right\rceil$}
\label{Fig:15}
\end{figure}

\begin{lemma}
The $\kappa_{0(5,5)}$ circuit $($Fig. \ref{Fig:15}$)$ implements the pair of 5-cycles $(2^{n-2}$ $-2$, $2^n-2$, $2^n-2^{n-2}-2$, $2^{n-1}-2$, $2^n-2^{n-3}-2)$ $(2^{n-2}-1$, $2^n-1$, $2^n-2^{n-2}-1$, $2^{n-1}-1$, $2^n-2^{n-3}-1)$ by cost $36n-206$.
\end{lemma}

\begin{pf}
It can be verified that the $\kappa_{0(5,5)}$ circuit shown in Fig. \ref{Fig:15}-a differs from the $\kappa_{0(5)}$ circuit in its least significant bit (i.e., the $0^{th}$ bit) which results in two 5-cycles. Applying Lemma 7.3 of \cite{Barenco95} on each gate shown in Fig. \ref{Fig:15}-a and canceling the resulted redundant gates transformed Fig. \ref{Fig:15}-a to Fig. \ref{Fig:15}-b. The total number of $36n-206$ elementary gates can be obtained by a summation of the costs of gates shown in Fig. \ref{Fig:15}-b.
\end{pf}

\begin{figure}[t]
\centering
\includegraphics[height=4cm]{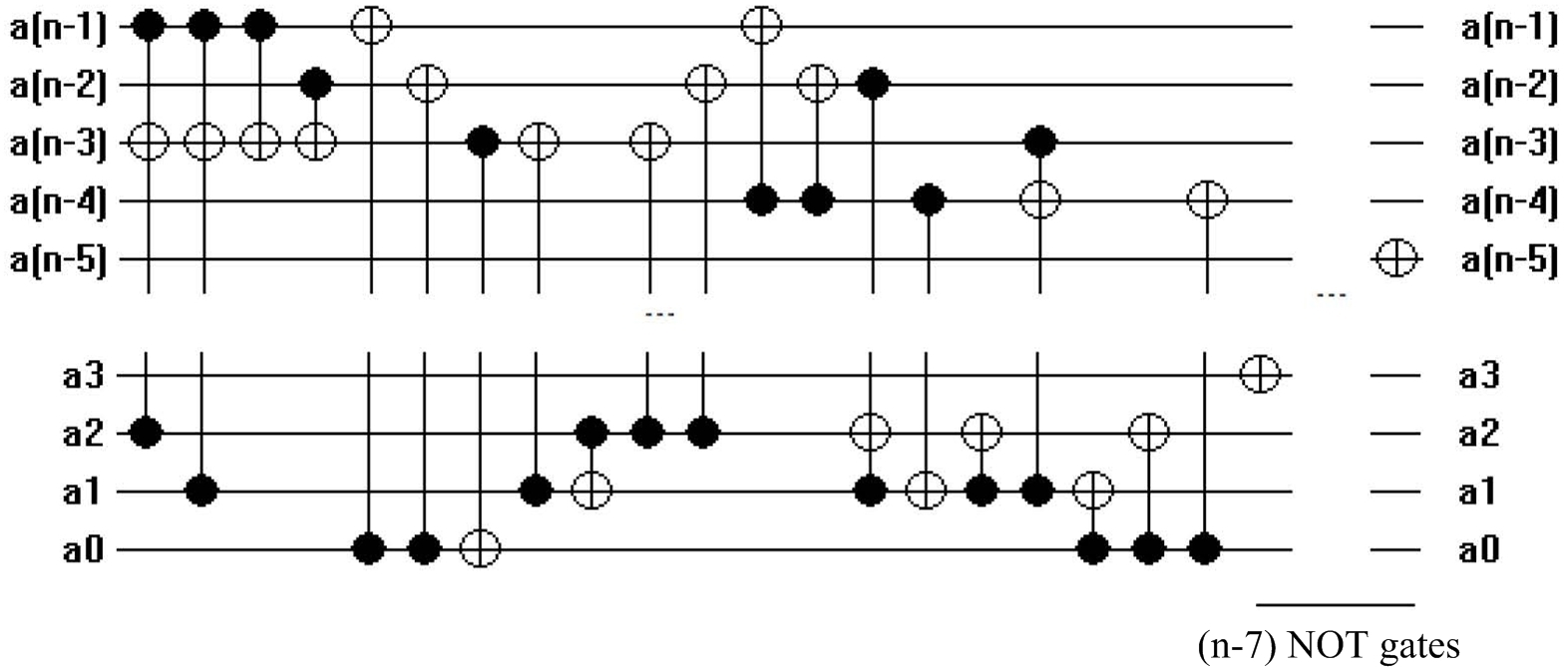}
\caption{The circuit of Theorem \ref{theorem:7}}
\label{Fig:16}
\end{figure}

\begin{theorem} \label{theorem:7}
$(Syn_{5,5}$ method$)$: An arbitrary 5-cycle $(a$, $b$, $c$, $d$, $e)$ $(f$, $g$, $h$, $i$, $j)$ can be implemented by at most $64n-54$ elementary gates.
\end {theorem}

\begin{pf}
Apply at most $13n+47$ elementary gates to convert $a$ to $2$ (i.e., $0 \cdots 0\textbf{1}0$), $d$ to $4$ (i.e., $0 \cdots 0\textbf{1}00$), $e$ to $2^{n-4}$ (i.e., $000\textbf{1}0 \cdots 0$), $f$ to $2^{n-3}$ (i.e., $00\textbf{1}0 \cdots 0$), $h$ to $2^{n-1}$ (i.e., $\textbf{1}0 \cdots 0$), $i$ to $2^{n-2}$ (i.e., $0\textbf{1}0 \cdots 0$), $j$ to $1$ (i.e., $0 \cdots 0\textbf{1}$), $b$ to $2^{n-1}+4$ (i.e., $\textbf{1}0 \cdots 0\textbf{1}00$), $c$ to $2^{n-1}+2$ (i.e., $\textbf{1}0 \cdots 0\textbf{1}0$) and $g$ to $2^{n-1}+2^{n-2}+2^{n-3}$ (i.e., $\textbf{11}10 \cdots 0$) sequentially. Note that $a$ and $d$ can be transformed to $2$ and $4$ by only CNOT gates, respectively. In addition, for each of other terms $e$, $f$, $h$, $i$, and $j$ at most one Toffoli gate and $n-1$ CNOT gates should be used. For the last three terms $b$, $c$, and $g$ at most two Toffoli gates should be used to set the control bits to $1$. Then, at most $n-2$ Toffoli gates should be applied for each term. For $n \geq 10$, the terms $b$, $c$, and $g$ can also be implemented by at most one Toffoli gate and $n-1$ CNOT gates which lead to $10n+32$ elementary gates. Now, apply the circuit shown in Fig. \ref{Fig:16}. By using at most $14n+76$ elementary gates, the terms $a$, $b$, $c$, $d$, $e$, $f$, $g$, $h$, $i$, and $j$ are changed to $2^{n-2}-2$, $2^n-2$, $2^n-2^{n-2}-2$, $2^{n-1}-2$, $2^n-2^{n-3}-2$, $2^{n-2}-1$, $2^n-1$, $2^n-2^{n-2}-1$, $2^{n-1}-1$, and $2^n-2^{n-3}-1$, respectively (the  $\pi_{(5,5)}$ circuit). Then, apply the $\kappa_{0(5,5)}$ and the $\pi_{(5,5)}^{-1}$ circuit to implement the cycles ($a$, $b$, $c$, $d$, $e$) ($f$, $g$, $h$, $i$, $j$) by at most $64n-54$ elementary gates.
\end{pf}


So far, direct implementations of the selected building blocks have been studied. Table \ref{Table:1} shows a summary of the achieved results for direct implementations of the selected building blocks. In this table, the maximum number of elementary gates of our direct synthesis method and the 2-cycle-based method \cite{Prasad06} for the set of proposed building blocks are compared. As demonstrated in this table, the direct $k$-cycle-based implementation has a significant potential to reduce the cost. However, as the direct implementation of a general $k$-cycle could be very hard, in this paper a decomposition algorithm is also proposed to be used in conjunction with the selected set of building blocks.

\begin{table}[tb]
	\caption{Maximum cost comparison for the proposed BBs}
	\centering		
	\begin{tabular}{lcccccc}
		\hline
		\multirow{2}{*}{\textbf{BB}} & \multirow{2}{*}{\textbf{Length}} & \multicolumn{4}{c}{\textbf{Our Approach}} & \textbf{\cite{Prasad06}} \\
		 & & $\kappa_0$ & $\pi$, $\pi^{-1}$ & Total & Cost/Length & Total \\ \hline 		
			(2,2) & 4 & 24\textit{n}-88 & 5\textit{n}+12 & 34\textit{n}-64 & 8.5\textit{n}-16 & 34\textit{n}-64 \\ \hline
			(3) & 3 & 24\textit{n}-88 & 4\textit{n}+3 & 32\textit{n}-82 & 10.7\textit{n}-27.3 & 68\textit{n}-128 \\ \hline
			(3,3) & 6 & 24\textit{n}-112 & 7\textit{n}+33 & 38\textit{n}-46 & 6.3\textit{n}-15.3 & 68\textit{n}-128 \\ \hline
			(2,4) & 6 & 36\textit{n}-180 & 7\textit{n}+29 & 50\textit{n}-122 & 8.3\textit{n}-20.3 & 68\textit{n}-128 \\ \hline
			(4,4) & 8 & 36\textit{n}-228 & 10\textit{n}+51 & 56\textit{n}-126 & 7\textit{n}-15.7 & 102\textit{n}-192 \\ \hline
			(5) & 5 & 48\textit{n}-166 & 6\textit{n}+18 & 60\textit{n}-130 & 12\textit{n}-26 & 102\textit{n}-192 \\ \hline
			(5,5) & 10 & 36\textit{n}-206 & 14\textit{n}+76 & 64\textit{n}-54 & 6.4\textit{n}-5.4 & 136\textit{n}-256 \\ \hline
			\end{tabular}
	\label{Table:1}
\end{table}

\subsection{Decomposition Method} \label{sec:decomposition}
In the rest of this paper, $2$, $3$, $4$ and $5$ cycles are called \emph{elementary cycles}. For an arbitrary single permutation $P$, we would like to decompose it into a set of elementary cycles like $c_1$, $c_2$, ..., $c_k$ such that applying $P$ would be identical to applying $c_1$, $c_2$, ..., $c_k$, sequentially; and $c_1$, $c_2$, ..., $c_k$ as well as $P$ would belong to a single permutation group.

To describe the decomposition method, the following notations are used: $P$ as an input permutation, $m$ as the maximum cycle length available in $P$, $C_k$ as a cycle of length $k$, $C_{k,i(k)}$ as the set of $i(k)$ cycles each of which is of length $k$, $C_k^j$ ($j \leq i(k)$) as the $j^{th}$ cycle of the cycle set $C_{k,i(k)}$, $N(k)$ as the number of disjoint 5-cycles in a given $k$-cycle, $L(k)$ as the length of a given $k$-cycle after detaching $N(k)$ disjoint 5-cycles, and $E(k)$ as the length of a given $k$-cycle after detaching all of the available disjoint/non-disjoint 5-cycles in the given $k$-cycle.

Any permutation $P$ can be written uniquely, except for the order, as a product of disjoint cycles. Without loss of generality, we assume that $P=C_{m,i(m)}$ $C_{m-1,i(m-1)}$ $\cdots$ $C_{3, i(3)}$ $C_{2,i(2)}$ where $\forall k \in$ ($2$, $\cdots$, $m$): $i(k)\geq 0$. For each $C_{k,i(k)}$ ($k>5$) in $P$, $C_{k,i(k)}$ is decomposed into a set of cycles of lengths $5$, $4$, $3$, and $2$, sequentially. In addition, for any two cycles $C_{k,i(k)}$ and $C_{j,i(j)}$ ($k>j$), $C_{k,i(k)}$ is processed first. Consider a given $k$-cycle ($1$, $2$, $3$, $4$, $\cdots$, $k$) ($k>5$). It is possible to decompose it into two cycles ($1$, $2$, $3$, $4$, $5$) ($6$, $7$, $\cdots$, $k$, $1$) of length 5 and ($k-4$), respectively. Repeating the process leads to $N(k)$= $\left\lfloor k/5 \right\rfloor$ disjoint 5-cycles and a cycle of length $L(k)$=$N(k)$+($k$ mod $5$) with some non-disjoint members. This process is called \emph{the 5-cycle extraction method} in the rest of the paper.

Since $C_{k,i(k)}$ $\forall k \in$ ($2$, $\cdots$, $m$) contains $i(k)$ cycles of length $k$, one can write $C_{k,i(k)}$= $C_k^1$ $C_k^2$ $\cdots$ $C_k^{i(k)}$. For each $C_k$ and by using the 5-cycle extraction method, $C_k$=$C_{5,1}$ $C_{k-4,1}$=$C_{5,2}$ $C_{k-8,1}$=...=$C_{5,N(k)}$ $C_{L(k),1}$. Repeating this process for $L(k)$, $L(L(k))$, etc. lead to $C_k$=$C_{5,N(k)}$ $C_{5,N(L(k))}$ $C_{5,N(L(L(k)))}$ $\cdots$ $C_{5,N(L(L \cdots (k)))}$ $C_{E(k),1}$. Note that $E(k)$ is smaller than 5. Since there are $i(k)$ cycles of length $k$, $C_{k,i(k)}$ = $C_{5,N(k) \times i(k)}$ $C_{5,N(L(k)) \times i(k)}$ $C_{5,N(L(L(k)))\times i(k)}$, ..., $C_{5,N(L(L \cdots (k)))\times i(k)}$ $C_{E(k),i(k)}$.

It can be verified that the resulted elementary cycle of a $k$-cycle ($k>5$) has no common members with other cycles. In addition, all disjoint/non-disjoint 5-cycles (detached from a $k$-cycle) are disjoint over other cycles. Therefore, the input permutation $P$ can be written as (\ref{eq:1}). See Example \ref{Example:2} for more details.

\begin{equation}
\label{eq:1}
\begin{array}{l}
 P = \left. {\left( {C_{5,N(k) \times i(k)} } \right)} \right|_{k = 5}^m \left. {\left( {C_{5,N(L(k)) \times i(k)} } \right)} \right|_{k = 5}^m ... \\
 \left. {\,\,\,\,\,\,\,\,\,\left( {C_{5,N(L(L(...(k))) \times i(k)} } \right)} \right|_{k = 5}^m C_{4,i'(4)} C_{3,i'(3)} C_{2,i'(2)}  \\
 where: \\
 i'(4) = i(4) + \sum\limits_{k = 5}^m {\left. {i(k)} \right|_{E(k) =  = 4} } , \\
 i'(3) = i(3) + \sum\limits_{k = 5}^m {\left. {i(k)} \right|_{E(k) =  = 3} } , \\
 i'(2) = i(2) + \sum\limits_{k = 5}^m {\left. {i(k)} \right|_{E(k) =  = 2} }  \\
 \end{array}
\end{equation}

\begin{example}\label{Example:2}
Consider $P=($3, 5, 6, 7, 9, 10, 11, 12, 13, 14, 15, 17, 18, 19, 20, 21$)$ $($22, 23, 24, 25, 26, 27$)$ $($28, 29$)$ $($30, 31$)$ written as $C_{16,i(16)}$ $C_{6,i(6)}$ $C_{2,i(2)}$. It can be verified that $m=16$, $i(16)=1$, $i(6)=1$, $i(2)=2$, and $i(k)=0$ for $k \in$ $(3$, $4$, $5$, $7$, $8$, $\cdots$, $15)$. We have:
\begin{itemize}
	\item $k=16$
			\begin{itemize}
				\item $N(16)$= $\left\lfloor 16/5 \right\rfloor$=$3$, $L(16)$=$3+1$=$4$
				\item $N(L(16))$=$N(4)$=$0$, $L(L(16))$=$4$=$E(16)$
			\end{itemize}
	\item $k=6$
			\begin{itemize}
				\item $N(6)$=$1$, $L(6)$=$2$=$E(6)$
			\end{itemize}
	\item $k=2$
			\begin{itemize}
				\item $N(2)$=$0$, $L(2)$=$2$=$E(2)$
			\end{itemize}
\end{itemize}

Therefore $P$= $(C_{5,3}$ $C_{5,1})$ $(C_{4,1}$ $C_{2,3})$ = {$(3, 5, 6, 7, 9)$ $(10, 11, 12, 13, 14)$ $(15, 17$, $18, 19, 20)$  $(22, 23, 24, 25, 26)$ $(21, 3, 10, 15)$ $(22, 27)$ $(28, 29)$ $(30, 31)$}.	
\end{example}

Considering the 5-cycle extraction method, the extraction time complexity of each $k$-cycle can be written as $O(k)$ + $O(L(k))$ + $O(L(L(k))$ + $\cdots$ + $O(E(k))$ $\leq O(\eta k)$ where $\eta$ is an integer smaller than $k$. Therefore, each given $k$-cycle is processed with the time complexity of $O(k)$. On the other hand, as there are $i(k)\geq 0$ cycles of length $k$, the total time complexity of the decomposition method is $O(m) \times i(m)$ + $O(m-1) \times i(m-1)$ + $\cdots$+ $O(2) \times i(2)$ where $O(i(k))$=$O(2^n/k)$, $k<2^n$ for $k \in (2 \cdots m)$. Therefore, we have $O(m) \times i(m)$ + $O(m-1) \times i(m-1)$ + $\cdots$+ $O(2) \times i(2)$=$O(m^2)$=$O(2^{2n})$ as $m<2^n$. It is important to note that the decomposition algorithm of \cite{Prasad06} works with the same $O(2^{2n})$ time complexity.
After the decomposition stage, the resulted elementary cycles should be implemented by using the proposed synthesis algorithms. Note that the total number of extracted 5-cycles is $O(k)$+$O(L(k))$+$O(L(L(k)))$+$\cdots$ which is equal to $O(k)$. Considering all $k$-cycles ($k \geq 5$), the total number of 5-cycles is $O(2^{2n})$ as explained above. In addition, as each $k$-cycle ($k \geq 5$) could produce at most one elementary cycle with length $2$, $3$ or $4$, the total number of elementary cycles is at most $\sum_{k=2 \cdots m} i(k)$=$O(2^n)$. Therefore, the total number of elementary cycles is $O(2^{2n})$ that leads to the time complexity of $O(2^{2n})\times O(Synthesis Algorithm)$. It can be verified that the proposed synthesis algorithms for the elementary cycles are of $O(n)$. As a result, the total time complexity of the proposed approach is $O(2^{2n} \times n)$, the same as \cite{Prasad06}.

To count the maximum number of elementary cycles in the proposed method, note that the number of 5-cycle pairs, 3-cycle pairs and 4-cycle pairs resulted from the decomposition algorithm are $Num_{5,5}$ = $\left\lfloor  \frac{{\rm 1}}{{\rm 2}} \sum_{k=5 \cdots m} {N(k)+N(L(k))+ \cdots}\right\rfloor$, $Num_{3,3}$ = $\frac{{\rm 1}}{{\rm 2}} i'(3)$, and $Num_{4,4}$ = $\frac{{\rm 1}}{{\rm 2}} i'(4)$, respectively. On the other hand, at most one single 5-cycle, one single 3-cycle and one 4-cycle followed by a 2-cycle are produced, i.e., $Num_5$ = mod($\sum_{k=5 \cdots m}$ {$N(k)$ + $N(L(k))$ + $\cdots$}, $2$), $Num_3$ = mod($i'(3), 2$) and $Num_{4,2}$ = mod($i'(4), 2$). Finally, the number of 2-cycle pairs is $Num_{2,2}$= $\left\lfloor \frac{{\rm 1}}{{\rm 2}} (i'(2)- Num_{4,2})\right\rfloor$. Altogether, the maximum number of elementary gates resulted in the proposed $k$-cycle-based synthesis method can be expressed by (\ref{eq:2}). See the following examples for more details.
\begin{equation}
\begin{array}{l}
 Num_{5,5}  \times ({\rm 6}4{\rm n - 54) } + Num_5  \times ({\rm 6}0{\rm n - 130)} + Num_{3,3}  \times ({\rm 3}8{\rm n - 46)} +  \\
 Num_3  \times ({\rm 3}2{\rm n - 82)} + Num_{4,4}  \times ({\rm 56n - 126)} + Num_{4,2}  \times ({\rm 50n - 122) } +  \\
 Num_{2,2}  \times ({\rm 34n - 64)} \\
 \end{array}
\label{eq:2}
\end{equation}

\begin{example} \label{Example:3}
Again, reconsider the permutation of Example \ref{Example:2}, $P$ = $C_{16,1}$ $C_{6,1}$ $C_{2,2}$ where $Num_{5,5}$ = $\frac{{\rm 1}}{{\rm 2}} \left\lfloor (N(16)+N(6))\right\rfloor$ = $2$, $Num_5$ = $0$, $Num_{3,3}$ = $0$, $Num_3$ = $0$, $Num_{4,4}$ = $0$, $Num_{4,2}$ = $1$, and $Num_{2,2}$ = $\left\lfloor \frac{{\rm 1}}{{\rm 2}} (3-1) \right\rfloor$ = $1$. At most $2 \times (64n-54)$ $+(50n-122)$ $+(34n-64)$ $=212n-294$ elementary gates are produced using our $k$-cycle-based synthesis method.
\end{example}

\begin{example} \label{Example:4}
Let $P$=$(3$, $5$, $6$, $7$, $9$, $10$, $11$, $12$, $13$, $14)$ $(15$, $17$, $18$, $19$, $20$, $21)$ $(22$, $23$, $24)$ $(25$, $26$, $27)$ $(28$, $29$, $30)$ ($P$ = $C_{10,1}$ $C_{6,1}$ $C_{3,3}$). After decomposition, we have $P$ $=$ $C_{5,3}$ $C_{3,3}$ $C_{2,2}$. After applying the proposed method, $Num_{5,5}$ $=$ $1$, $Num_5$ $=$ $1$, $Num_{3,3}$ $=$ $1$, $Num_3$ $=$ $1$, $Num_{4,4}$ $=$ $0$, $Num_{4,2}$ $=$ $0$, $Num_{2,2}$ $=$ $1$ and at most $2 \times (64n-54)$ $+$ $(60n-130)$ $+$ $(38n-46)$ $+$ $(32n-82)$ $+$ $(34n-64)$ $=$ $292n-430$ elementary gates are produced.
\end{example}

\begin{figure}[t]
\centering

\begin{quote}
\begin{tabbing}
{\bf Step1} \\
Fix 0 and $2^i$ terms use a pre-process stage as done in \cite{Shende03}.\\\\
{\bf Step2} \\
if $n < 7$\\
\hspace*{1em} 1- Decompose the input permutation into a set of 2-cycles.\\
\hspace*{1em} 2- Apply $Syn_{2,2}$ to synthesize all 2-cycles\\
else\\
\hspace*{1em}\= 1- Decompose the input permutation into a set of 5, 4, 3, and 2 cycles \\
\hspace*{1em}\= 2- Synthesize all disjoint 5-cycle pairs ($Syn_{5,5}$)\\
\hspace*{1em}\= 3- Synthesize single 5-cycles ($Syn_{5}$)\\
\hspace*{1em}\= 4- Synthesize all disjoint 3-cycle pairs ($Syn_{3,3}$)\\
\hspace*{1em}\= 5- Synthesize single 3-cycles ($Syn_{3}$)\\
\hspace*{1em}\= 6- Synthesize all disjoint 4-cycle pairs ($Syn_{4,4}$)\\
\hspace*{1em}\= 7- Synthesize all disjoint 4-cycle and 2-cycle pairs ($Syn_{4,2}$)\\
\hspace*{1em}\= 8- Synthesize all disjoint 2-cycle pairs ($Syn_{2,2}$)\\

\end{tabbing}
\end{quote}

\caption{The $k$-cycle-based synthesis method}
\label{Fig:17}
\end{figure}

As stated at the beginning of Section \ref{sec:our_method}, the zero and $2^i$ terms are fixed by applying a few Toffoli and CNOT gates as done in \cite{Shende03}. In addition, For small $n$ (i.e., $n < 7$), the decomposition algorithm is modified to produce only 2-cycles where each cycle pair is synthesized by the $Syn_{2,2}$ method. The complete $k$-cycle-based synthesis method is shown in Fig. \ref{Fig:17}. For $n \geq  7$, a given permutation is recursively decomposed into a set of elementary cycles each of which is synthesized by the synthesis algorithm listed in parentheses as discussed.

\begin{theorem} \label{corollary:1}
The proposed $k$-cycle-based synthesis method always converges.

\end{theorem}

\begin{pf}
According to the proofs of Theorem \ref{theorem:1} to Theorem \ref{theorem:7}, the suggested building blocks (i.e., a pair of 2-cycles, single 3-cycle, a pair of 3-cycles, single 5-cycle, a pair of 5-cycles, a single 2-cycle (4-cycle) followed by a single 4-cycle (2-cycle), and a pair of 4-cycles) can always be synthesized for any arbitrary values of cycle elements for $n \geq 7$ as far as each cycle element is neither 0 nor $2^i$. In addition, by using the proposed decomposition algorithm, a given large cycle can always be decomposed into a set of elementary cycles. For small $n$ (i.e., $n < 7$), the decomposition algorithm produces only 2-cycles where each pair can always be synthesized by the $Syn_{2,2}$ method. Considering the pre-process stage for the zero and $2^i$ terms and the synthesis scenarios for $n < 7$ and $n \geq 7$ as explained above lead to the theorem.
\end{pf}
\subsection{Worst Case Analysis} \label{sec:worst_case}

To analyze the total number of elementary gates resulted from the proposed $k$-cycle-based synthesis method in the worst case, assume that the maximum of $m$ members ($a_1$, $a_2$, $\cdots$, $a_m$) of a given permutation $P$ are moved. As each $a_k$, $k \in$ ($2, \cdots, m$) is neither $0$ nor $2^i$, $m$ is equal to $2^n-n-1$ for an even $n$ and equal to $2^n-n-2$ for an odd $n$.

\begin{theorem} \label{theorem:8}
The maximum number of elementary gates in the proposed cycle-based synthesis method is calculated by $8.5n2^n + o(2^n)$.
\end{theorem}

\begin{pf}
In order to place each row at its right position, several reversible gates should be applied in the proposed method. The worst-case cost occurs for the maximum number of changed rows (i.e., $m=o(2^n)$). The synthesis costs listed in Table \ref{Table:1} (i.e., Cost/Length) indicate that the cost of correcting a single row is $8.5n-16$ for a pair of 2-cycles, $10.7n-27.3$ for a single 3-cycle, $6.3n-15.3$ for a pair of 3-cycles, $8.3n-20.3$ for a single 2-cycle followed by a single 4-cycle, $7n-15.7$ for a pair of 4-cycles, $12n-26$ for a single 5-cycle and $6.4n-5.4$ for a pair of 5-cycles.

For a decomposition with $2^n$ changed rows, there are at most one single 5-cycle and one single 3-cycle. Considering the cost of $12n-26$ for correcting a single row in a single 5-cycle, $10.7n-27.3$ in a single 3-cycle and $8.5n-16$ in a pair of 2-cycles, it can be verified that the worst-case cost for a decomposition with $2^n$ changed rows is $8.5n2^n+o(2^n)$.
\end{pf}

Theorem \ref{theorem:8} shows a lower upper bound for $k$-cycle-based synthesis method compared to the best reported upper bound of $11n2^n+o(n2^n)$ for the synthesis algorithm proposed in \cite{MaslovTODAES07}. Given the fact that the $8.5n2^n$ term is dominant over the $o(2^n)$ term, the former will be used in the remainder of this subsection for cost analysis.

Reversible logic has application in quantum computing \cite{Barenco95}, \cite{Nielsen00}. Most quantum algorithms presume that interaction between arbitrary qubits is possible with no extra cost. However, some restrictions exist in real quantum technologies \cite{Shende06}. For example in a \emph{Linear Nearest Neighbor (LNN)} architecture, only adjacent qubits may interact. The implementation complexity with limited interaction depends on the relative target and control positions. It can be modeled by using a sequence of SWAP gates to move controls and targets close to each other to construct appropriate gates. Theorem \ref{theorem:10} examines the proposed method for LNN architecture.

\begin{theorem} \label{theorem:10}
The maximum number of elementary gates in the proposed $k$-cycle-based synthesis method for LNN architecture is equal to $51n^22^n$.
\end{theorem}

\begin{pf}
To prove, the number of required SWAP operations performing a 2-qubit gate $g$ with control $c$ and target $t$ has to be found. We assume $c>t$. It can be verified that ($c-t-1$) SWAP operations are required to bring the control adjacent to the target, one gate is required to perform $g$, and the same sequence of ($c-t-1$) SWAP operations are required to return value of the $i^{th}$ ($t<i\leq c$) qubit to its initial value. Considering a cost 3 for each SWAP operation leads to $6 \times (c-t-1)+1$. The case of $c\leq t$ can be readily deduced by following the same approach.

The theorem can be proven by using Theorem \ref{theorem:8} and plugging in the cost found above.
\end{pf}

\section {Experimental Results} \label{sec:exper}

The proposed $k$-cycle-based synthesis method and the 2-cycle-based algorithm presented in \cite{Prasad06} were implemented in C++ and all of the experiments were done on an Intel Pentium IV 2.2GHz computer with 2GB memory. In addition, we used one of the most recent and efficient NCT-based synthesis tools proposed in \cite{MaslovTODAES07} for our comparisons. This method used Reed-Muller (RM) spectra in an iterative synthesis procedure (RM-based method). In all experiments, the post-processing algorithm proposed in \cite{Prasad06} was applied to simplify circuits produced by our synthesis method and the algorithm of \cite{Prasad06}. \label{Reviwer:14} In this method, optimal circuits for all 40320 3-input reversible functions and a large set of 4-input circuits were generated and stored in a compact data-structure. As a result, applying the post-processing algorithm of \cite{Prasad06} leads to optimal results for all 3- and some 4-input specifications. The synthesis algorithm of \cite{MaslovTODAES07} was applied in \emph{\textquotedblleft synthesized/ resynthesized using 3 methods\textquotedblright} mode for circuits with $n<15$ and in \emph{\textquotedblleft synth/resynth with MMD} (\emph{15+ variables})\textquotedblright for $n\geq15$. In addition, the synthesis algorithm, the template matching method, the random and exhaustive driver algorithms were applied sequentially to synthesize each function with a time limit of 12 hours as done in \cite{MaslovTODAES07}. Bidirectional and quantum cost reduction modes were also applied.

To evaluate the proposed synthesis method, the completely specified reversible benchmarks from \cite{MaslovSite} were examined. In addition, the best documented synthesis costs available at \cite{MaslovSite} resulted from applying different NCT-based synthesis tools were used for our comparisons. In some cases, the synthesis results in NCT library for some benchmarks have not been reported yet (these functions are N-th prime functions over more than 7 bits, hamming coding functions (hwb) over more than 11 bits\footnote{For hwb functions, polynomial size reversible circuits in NCTF library (NCT library plus the Fredkin gate \cite{Fredkin82}) with $[log(n)]+1$ garbage bits and $O(n log^2(n))$ gates exist \cite{MaslovSite}.} and permanent functions). In those cases, we applied the synthesis method of \cite{MaslovTODAES07} which works efficiently in terms of quantum cost with a time limit of 12 hours. If it failed to synthesize a function in the given time limit (for hwb functions over more than 11 bits and N-th prime functions over more than 10 bits, the algorithm failed), the method of \cite{Prasad06} was applied.
All synthesis algorithms were compared in terms of the quantum cost as done in \cite{MaslovSite}. Our actual circuits are available from \cite{SaeediSite}.

The results of the proposed $k$-cycle-based synthesis method (\textsc{Pure $k$-cycle}) and the best synthesized circuits resulted from the previous NCT-based synthesis algorithms (\textsc{Best Results}) were shown in Table \ref{table:2}.
A comparison of the synthesis costs of the proposed $k$-cycle-based method and the best reported ones reveals that the cycle-based approach treats differently in terms of the quantum cost for different benchmarks (for examples see the results of hwb11 and cycle10\_2). In the rest of this section, by analyzing the characteristics of different benchmarks, a hybrid synthesis framework is proposed which uses the cycle-based method in conjunction with the method of \cite{MaslovTODAES07} to synthesize a given function. As shown later, the proposed hybrid framework can improve the average quantum costs efficiently.

{\setlength{\tabcolsep}{3pt}
\begin{table}[!ht]
\caption{The comparison costs of the proposed synthesis framework. Time values are in seconds.}
\label{table:2}
\scriptsize
\centering
\begin{tabular}{|c|l|c|c|c|c|c|c|c|c|}
\hline
\multirow{3}{*}{\textsc{Cat.}} & \multirow{3}{*}{\textsc{Benchmark}}  & \multirow{3}{*}{\textsc{\textit{n}}} & \multirow{3}{*}{\textsc{Dist.}}&  & \textsc{Pure} & \multicolumn{3}{|c|}{\textsc{The Proposed}}  & \textsc{Cost}\\

& &  & & \textsc{Best Results} & \textsc{$k$-Cycle} & \multicolumn{3}{|c|}{\textsc{Hybrid Framework}}  & \textsc{Impr.}\\

&	\textsc{Function}	&  	& 	& Cost & Cost & Cost & Time & {\textsc{Method}} & (\%)\\

\hline
\multirow{12}{*}{1}&3\_17	&3	  & 0.18 	& 12	&  12 & 12
& 4 & kC+R & 0\\
\cline{2-10}

&4\_49		&4	 & 0.37 	& 32	& 116 & 32
& 5 & kC+R & 0\\
\cline{2-10}

&ham3		&3   & 0.06& 7	&  7 & 7
& 4 & kC+R & 0\\
\cline{2-10}

&hwb4		&4  & 0.36 	& 23	& 60 & 24
& 30 & kC+R & -4\\
\cline{2-10}

&hwb5		&5   & 0.44 	&104	& 196 & 91
& 32 & kC+R & 13 \\
\cline{2-10}

&hwb6		&6	   & 0.49 	&140	& 526 & 107
 & 44 & kC+R & 24\\
\cline{2-10}

&mod5adder	& 6	  & 0.07 	&	77 & 853 & 79
  & 20 & kC+R & -3\\
\cline{2-10}

&nth\_prime3\_inc	&	3  &  0.13&	6	& 6 & 6
  & 3 &  kC+R & 0\\
\cline{2-10}

&nth\_prime4\_inc	&	4  & 0.47 &	58	& 190 & 51
  & 20 & kC+R & 12\\
\cline{2-10}

&nth\_prime5\_inc	&	5  & 0.34	&	91	& 363 & 97
  & 27 &  kC+R & -7\\
\cline{2-10}

&nth\_prime6\_inc	&	6  &  0.61 & 	667	& 1314 & 701
  & 37 &  kC+R & -5\\
\cline{2-10}

&permanent2x2  &	6   & 0.02 	&	47& 227 & 49
  & 20  & kC+R & -4\\
\hline
\hline
\multicolumn{9}{|l|}{\textsc{average}} & 2\\
\hline
\hline

\multirow{22}{*}{2}&hwb7		&7	   & 0.54 	& 2611	& 2630 & 2630 
& 111 & kC & -1\\
\cline{2-10}

&hwb8	&8	   & 0.58&7013	& 6940 & 6940 
& 56  & kC & 1\\
\cline{2-10}

&hwb9	&9	 & 0.60	&22502	& 16173 & 16173 
 & 44 & kC & 28\\
\cline{2-10}

&hwb10	&10	   & 0.62	&59191	& 35618 & 35618 
 & 50 &  kC & 40\\
\cline{2-10}

&hwb11	&11	  &  0.63	&136756	& 90745 & 90745 
 & 60 & kC & 34\\
\cline{2-10}

&hwb12	&12	  & 0.64  	&	334218 & 198928 & 198928 
  &  122 & kC & 40\\
\cline{2-10}

&hwb13	&13	  &0.66 	&	935322 & 436305 & 436305 
  & 481  &  kC& 53\\
\cline{2-10}

&hwb14	&14	  & 0.65 & 	1818773 & 994340 & 994340 
  & 994 & kC & 45\\
\cline{2-10}

&hwb15	&15	  & 0.66 & 	4119568 & 1999194 & 1999194 
  & 1503 &  kC& 51 \\ 
\cline{2-10}

&hwb16	&16	  & 0.66 &	8910859 & 4730024 & 4730024
 & 4312 &kC& 47\\
\cline{2-10}

&nth\_prime7\_inc 	&	7  &  0.59 	&	2695	& 3172 & 3172 
  & 41 & kC & -18 \\
\cline{2-10}

&nth\_prime8\_inc 	&	8   & 0.62 	&	9409	&7618  & 7618
  & 56 & kC & 19\\
\cline{2-10}

&nth\_prime9\_inc 	&	9  &  0.55	&20888	& 17975 & 17975 
  & 60 & kC & 14\\
\cline{2-10}

&nth\_prime10\_inc 	&	10  &0.64  	&	48435	&40299  & 40299 
  & 64 & kC & 17\\
\cline{2-10}

&nth\_prime11\_inc 	&	11  &  0.62& 		197606	&95431 & 95431 
  & 89 & kC & 52\\
\cline{2-10}

&nth\_prime12\_inc 	&	12  & 0.61 & 		452301	& 208227& 208227 
  & 190 &  kC& 54\\
\cline{2-10}

&nth\_prime13\_inc 	&	13  & 0.6 & 		1016567	&474660 & 474660 
  & 420 & kC & 53\\
\cline{2-10}

&nth\_prime14\_inc 	&	14  &  0.62 & 	2254198	&1018661 & 1018661 
  & 1101 &  kC& 55\\
\cline{2-10}

&nth\_prime15\_inc 	&	15  &  0.63 & 	4948477	& 2271370& 2271370 
  & 2812 & kC &  54 \\
\cline{2-10}

&nth\_prime16\_inc 	&	16  &  0.64 &		10786095	&4823320 & 4823320 
  & 4018 & kC & 55\\ 
\cline{2-10}

&nth\_prime17\_inc 	&	17  &  0.61 & 	22144391 & 10592640 & 10592640 
  & 9231 & kC & 52\\ 
\hline
\hline
\multicolumn{9}{|l|}{\textsc{average}} & 35 \\
\hline
\hline

\multirow{6}{*}{3}&ham7		&7   & 0.38	&  49	& 2117 & 49
& $*$ & RM & 0\\
\cline{2-10}

&ham15		&15	   & 0.31& 214	& 140343 & 214	
& $*$ & RM & 0 \\
\cline{2-10}

&mod1024adder	& 20	  & 0.66& 1575	& 110222 & 1575
& $*$ & RM & 0\\
\cline{2-10}

&cycle10\_2	&12	 & 0.001 &  1206	& 93086 & 1206
 & $*$ & RM & 0 \\
\cline{2-10}

&cycle17\_3	&20	  & $\approx0$	& 6057	& 523891 & 6057
 & $*$  & RM & 0\\
\cline{2-10}

&permanent3x3  & 12	  &  $\approx0$	&	1884& 89777 & 1884
  & * & RM & 0\\
\hline
\hline
\multicolumn{9}{|l|}{\textsc{average}} & 0 \\
\hline

\end{tabular}
\begin{flushleft}
$*$ A time limit of 12 hours was considered in applying the method of \cite{MaslovTODAES07}.\\
\end{flushleft}
\end{table}
}

\begin{figure}[t]
\centering
\includegraphics[height=3cm]{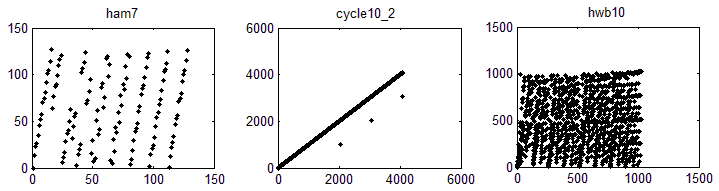}
\caption{The distributions of output code words for three benchmark functions}
\label{Fig:18}
\end{figure}

To evaluate the behavior of $k$-cycle-based synthesis method, a \emph{Distance} metric is defined as (\ref{eq:distance}) for each reversible function $f$ where $0\leq Distance(f) \leq 1$.
\begin{equation}\label{eq:distance}
    Distance(f)={\sum_{i=0}^{i=2^n-1}|f(i)-i|}/(2^{2n-1})
\end{equation}
For a given function $f$, \emph{Distance(f)} models the distribution of output code words compared with the identity function. Fig. \ref{Fig:18} shows the distributions of output code words for three benchmarks. As illustrated in this figure, ham7 ($Distance(f)=0.38$) and cycle10\_2 ($Distance(f)=0.001$) are more similar to the identity function ($f(i)=i, Distance(f)=0$) compared with hwb10 ($Distance(f)=0.62$). The distributions of output code words for other functions were reported in Table \ref{table:2} (i.e., \textsc{Dist.}).

\begin{figure}[t]
\centering
\includegraphics[height=8cm]{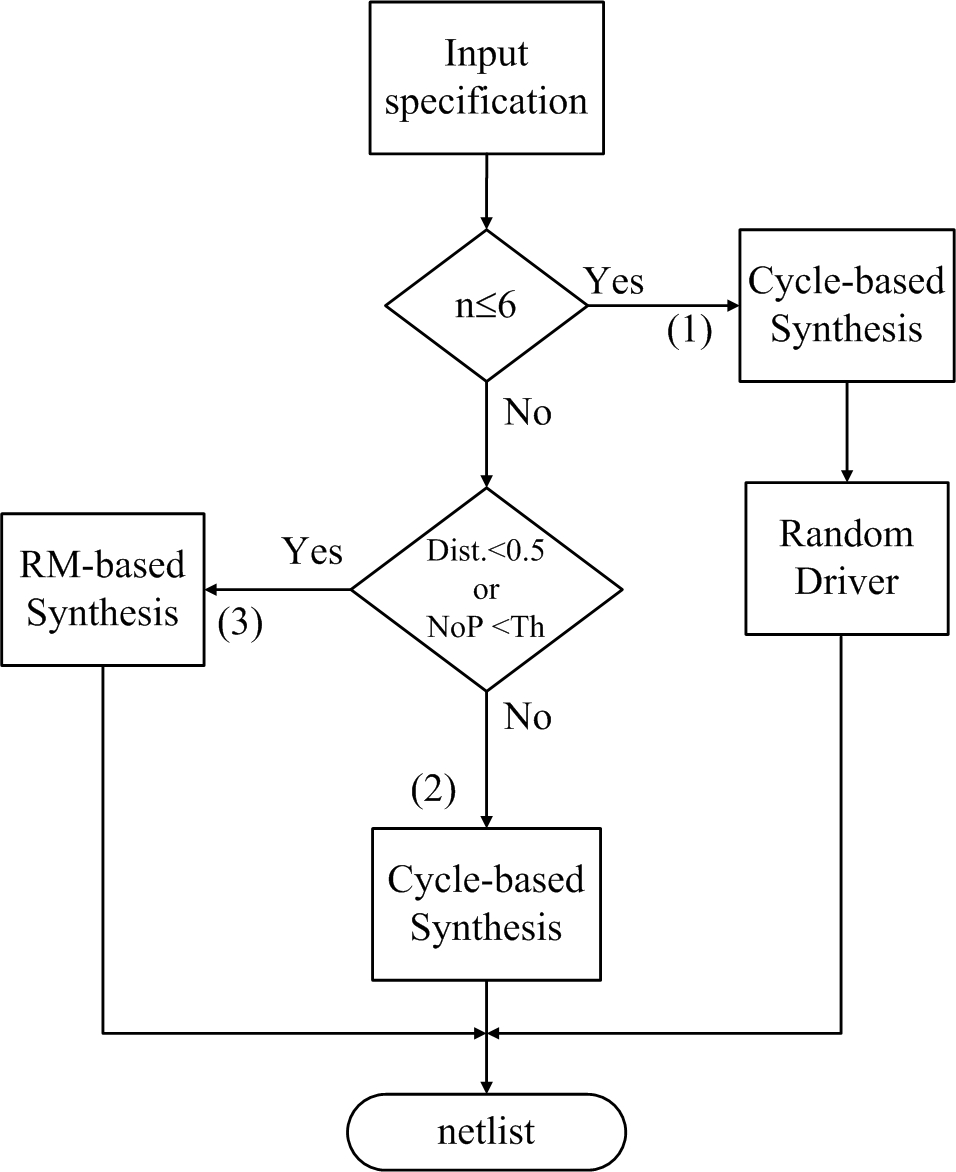}
\caption{The hybrid synthesis framework}
\label{Fig:flow}
\end{figure}

Based on the characterization of a reversible function, we divided benchmarks into three categories as shown in Table \ref{table:2} (\textsc{Cat.}). Category 1 includes small functions with less than seven inputs. Category 2 and category 3 include large functions with $n \geq 7$ but with different distribution levels. In other words, for each function in category 2 (3), $Distance(f)$ is greater (less) than $0.5$. By applying a hybrid synthesis framework, functions in different categories are handled differently as shown in Fig. \ref{Fig:flow}.

For functions in category 1, we applied the cycle-based synthesis method first. Then, the random\_driver procedure introduced in \cite{MaslovTODAES07} was applied.
Since category 1 includes small functions, applying the random\_driver method for optimizing the results has no runtime overhead. Hence, combining different heuristics (i.e., cycle-based approach and random\_driver procedure) to achieve better cost is reasonable. On the other hand, for large functions in category 2 with considerable differences from the identity function ($Distance \geq 0.5$), only the cycle-based synthesis method was applied. According to \cite{MaslovTODAES07}, for some functions in this category (i.e., hwb11) the method of \cite{MaslovTODAES07} needs several hours to synthesize the function. Similarly, in \cite{Donald08}, the authors stated their synthesis algorithm cannot synthesize hwb circuit with over five variables by NCT library (with 4GB RAM and finite runtime). Memory/runtime limitations will be even more challenging for hwb functions with more variables. As can be seen in Table \ref{table:2}, both average cost and runtime were improved for functions in category 2.

\begin{figure}[t]
\centering
\includegraphics[height=3cm]{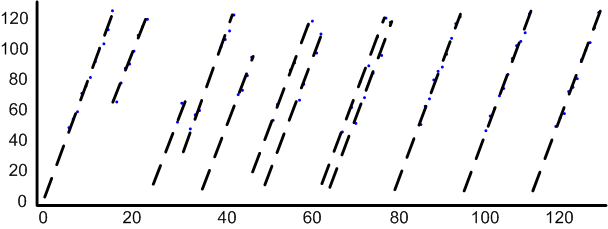}
\caption{Output patterns for ham7 function with $NoP=12$}
\label{Fig:19}
\end{figure}

On the other hand, for functions in category 3 which have some similarities to the identity function ($Distance < 0.5$), RM-based method is used in the proposed hybrid framework. A reversible function with large $Distance$ can have regular distribution at its output side (e.g., $f(i)=2^{n-1}-i$ where $Distance(f)=1$). Hence, number of patterns (NoP) in the distribution of output code words was also used in the proposed hybrid framework. Regular output distribution leads to a small NoP. Fig. \ref{Fig:19} shows output patterns for ham7 function ($NoP=12$). A function with an appropriate number of patterns ($NoP<Th$) at its output code words is similar to the identity function to some extent. Hence, such function was synthesized by using the RM-based method too. For example, mod1024adder with $Distance=0.66$ and $NoP=1000$ was synthesized by applying the RM-based method. We set $Th=0.005 \times 2^n$ in our experiments.

The results of hybrid synthesis framework were shown in Table \ref{table:2} where $k$-cycle-based, random\_driver and RM-based methods were denoted by kC, R, and RM, respectively. Runtime results (in seconds) for the hybrid framework were reported in Table \ref{table:2} too. According to the experimental results, RM-based method works very fast for functions in category 3 compared with category 2. Therefore, the proposed hybrid synthesis framework outperforms the best results in terms of quantum cost and runtime on average. Our synthesis tool potentially can synthesize functions with any number of variables. However, as the number of variables and resulted synthesized gates grows, the runtime and memory usage grow accordingly (for hwb functions with $n \geq 20 $, peak memory usage was more than 2GB).

Since both cycle-based and RM-based methods \cite{MaslovTODAES07} always result in a synthesized circuit, the proposed framework always converges. Moreover, as a generic reversible function $f$ with large $n$ and $Distance(f) \geq 0.5$ without regular patterns at its output side needs much more gates in the proposed hybrid framework compared with other functions, the worst-case cost of hybrid framework is identical to the worst-case cost of the cycle-based method (i.e., $8.5n2^n+o(2^n)$).

\section{Conclusion and future directions} \label{sec:conclusion}
In this paper, a $k$-cycle-based synthesis method for reversible functions was proposed and analyzed in detail. To this end, a set of synthesis algorithms was proposed to synthesize cycles of length less than 6 (i.e., elementary cycles). In addition, a decomposition algorithm was introduced to decompose a large cycle into a set of elementary cycles. Next, the decomposition algorithm and the proposed synthesis algorithms were used to synthesize all permutations. By evaluating different benchmark functions, the behavior of cycle-based synthesis method was analyzed and a hybrid synthesis framework was introduced which uses the proposed cycle-based synthesis method in conjunction with one of the recent synthesis methods.
\\
Our worst-case analysis revealed that the proposed hybrid synthesis framework leads to a lower upper bound compared to the present synthesis algorithms. The hybrid framework always converges and it leads to better average runtime. The experiments for average-case costs revealed that the proposed framework produces circuits with lower costs for benchmark functions.
\\
A natural next step to continue this path is working on the synthesis of cycles with length greater than 5 for the average-case cost improvement in the $k$-cycle-based synthesis method which can improve the results of hybrid framework too. In addition, working on a synthesis approach for incompletely specified functions based on the one proposed here could be considered as a future research.
\section*{Acknowledgment}
We would like to acknowledge Dmitri Maslov from University of Waterloo for providing an executable version of his recent synthesis tool.



\begin{thebibliography}{10}

\bibitem{Landauer61}
R.~Landauer.
\newblock Irreversibility and heat generation in the computing process.
\newblock {\em IBM Journal of Research and Development}, 5:183--191, July 1961.

\bibitem{Bennett73}
C.~Bennett.
\newblock Logical reversibility of computation.
\newblock {\em IBM Journal of Research and Development}, 17(6):525--532,
  November 1973.

\bibitem{Zhirnov03}
V.~V. Zhirnov, R.~K. Kavin, J.~A. Hutchby, and G.~I. Bourianoff.
\newblock Limits to binary logic switch scaling - a gedanken model.
\newblock {\em Proceedings of the IEEE}, 91(11):1934--1939, 2003.

\bibitem{Schrom98}
G.~Schrom.
\newblock {\em Ultra-Low-Power CMOS Technology}.
\newblock PhD thesis, Technischen Universitat Wien, June 1998.

\bibitem{Nielsen00}
M.~Nielsen and I.~Chuang.
\newblock {\em Quantum Computation and Quantum Information}.
\newblock Cambridge University Press, New York, 2000.

\bibitem{MaslovTODAES07}
D.~Maslov, G.~W. Dueck, and D.~M. Miller.
\newblock Techniques for the synthesis of reversible toffoli networks.
\newblock {\em ACM Trans. Des. Autom. Electron. Syst.}, 12(4):42, 2007.

\bibitem{MaslovTCAD08}
D.~Maslov, G.~W. Dueck, D.~M. Miller, and C.~Negrevergne.
\newblock Quantum circuit simplification and level compaction.
\newblock {\em IEEE Trans. on CAD}, 27(3):436--444, March 2008.

\bibitem{Gupta06}
P.~Gupta, A.~Agrawal, and N.K. Jha.
\newblock An algorithm for synthesis of reversible logic circuits.
\newblock {\em IEEE Trans. on CAD}, 25(11):2317--2330, 2006.

\bibitem{Shende03}
V.~V. Shende, A.~K. Prasad, I.~L. Markov, and J.~P. Hayes.
\newblock Synthesis of reversible logic circuits.
\newblock {\em IEEE Trans. on CAD}, 22(6):710--722, June 2003.

\bibitem{Barenco95}
A.~Barenco, C.~Bennett, R.~Cleve, D.~DiVincenzo, N.~Margolus, P.~Shor,
  T.~Sleator, J.~Smolin, and H.~Weinfurter.
\newblock Elementary gates for quantum computation.
\newblock {\em APS Physical Review A}, 52:3457--3467, 1995.

\bibitem{Negrevergne06}
C.~Negrevergne, T.~S. Mahesh, C.~A. Ryan, M.~Ditty, F.~Cyr-Racine, W.~Power,
  N.~Boulant, T.~Havel, D.~G. Cory, and R.~Laflamme.
\newblock Benchmarking quantum control methods on a 12-qubit system.
\newblock {\em Physical Review Letters}, 96(17), 2006.

\bibitem{MaslovTCAD05}
D.~Maslov, G.~W. Dueck, and D.~M. Miller.
\newblock Toffoli network synthesis with templates.
\newblock {\em IEEE Trans. on CAD}, 24(6):807--817, 2005.

\bibitem{Prasad06}
Aditya~K. Prasad, Vivek~V. Shende, Igor~L. Markov, John~P. Hayes, and Ketan~N.
  Patel.
\newblock Data structures and algorithms for simplifying reversible circuits.
\newblock {\em J. Emerg. Technol. Comput. Syst.}, 2(4):277--293, 2006.

\bibitem{MaslovDATE05}
D.~Maslov, C.~Young, D.~M. Miller, and G.~W. Dueck.
\newblock Quantum circuit simplification using templates.
\newblock In {\em DATE '05: Proceedings of the conference on Design, Automation
  and Test in Europe}, pages 1208--1213, Washington, DC, USA, 2005. IEEE
  Computer Society.

\bibitem{Shende06}
V.~V. Shende, S.~S. Bullock, and I.~L. Markov.
\newblock Synthesis of quantum-logic circuits.
\newblock {\em IEEE Trans. on CAD}, 25(6):1000--1010, June 2006.

\bibitem{MaslovSite}
D.~Maslov, G.~Dueck, and N.~Scott.
\newblock Reversible logic synthesis benchmarks page.
\newblock {\em http://www.cs.uvic.ca/\textasciitilde dmaslov/}, November 2009.

\bibitem{Fredkin82}
E.~F. Fredkin and T.~Toffoli.
\newblock Conservative logic.
\newblock {\em International Journal of Theoretical Physics}, 21(3/4):219--253,
  1982.

\bibitem{SaeediSite}
M.~Saeedi, M.~Saheb Zamani, and M.~Sedighi.
\newblock Reversible logic synthesis benchmarks.
\newblock {\em http://ceit.aut.ac.ir/QDA/benchmarks}, March 2010.

\bibitem{Donald08}
James Donald and Niraj~K. Jha.
\newblock Reversible logic synthesis with fredkin and peres gates.
\newblock {\em J. Emerg. Technol. Comput. Syst.}, 4(1):1--19, 2008.

\end{thebibliography}
\end{document}